\documentclass[aps,prc,showpacs]{revtex4} 

\let\revappendix\appendix
\usepackage[utf8]{inputenc}
\usepackage[english]{babel}
\usepackage{lipsum}
\usepackage{multirow}
\usepackage{soul}
\usepackage{amsmath} 
\usepackage{amssymb} 
\usepackage{graphics}
\usepackage{epsfig}
\usepackage{graphicx}
\usepackage{xfrac}
\usepackage{relsize}
\usepackage{slashed}
\newcommand{\mean}[1]{\left\langle #1\right\rangle}

\newcommand\alb{\begin{align}}
\newcommand\ale{\end{align}}

\def\be		{\begin{eqnarray}}
\def\en		{\end{eqnarray}}
\def\nen	{\nonumber\end{eqnarray}}
\def\no		{\nonumber}
\def\la		{\left\langle}
\def\ra		{\right\rangle}
\def\lt		{\left(}
\def\rt		{\right)}
\def\lq		{\left[}
\def\rq		{\right]}

\def\ds		{\ensuremath{\displaystyle}}
\def\hh		{\hspace{5 mm}}
\def\ug		{\ensuremath{\!&=&\!}}

\def\ov		{\ensuremath{\overline}}

\def\e		{\ensuremath{\epsilon}}
\def\N 		{\ensuremath{\mathbb{N}}}

\def\rr		{\ensuremath{\left\langle r^2\right\rangle_E^{1/2}}}

\def\R        {\ensuremath{\mathbb{R}}}
\def\N        {\ensuremath{\mathbb{N}}}

\def\bmi		 {\begin{minipage}}
\def\emi		 {\end{minipage}}
\def\wg   	{\ensuremath{{G}}}

\usepackage{longtable}
\usepackage{latexsym}

\usepackage{amssymb}
\usepackage{amsfonts} 
\usepackage{amsmath}
\begin{document}
\title{Sensitivity of the elastic electron-proton cross section to the proton radius}
\author{Simone~Pacetti  } 
\affiliation{\it         Dipartimento di Fisica e Geologia, and INFN Sezione di Perugia, 06123 Perugia, Italy}

\author{Egle~Tomasi-Gustafsson} 
\affiliation{\it IRFU, CEA, Universit\'e Paris-Saclay, 91191 Gif-sur-Yvette, France }

\date{\today}
\begin{abstract}
The precise determination of the proton radius from recent elastic scattering electron-proton data is discussed. The necessary precision on the elastic cross section to discriminate among the values coming from atomic spectroscopy is scrutinized in terms of the relevant quantity, i.e., the derivative of the form factor. It is shown that such precision is two orders of magnitude higher than the precision on the cross section, that is the  measured  observable. Different fits on the available data and of their discrete derivative, with analytical constraints are shown. The systematic error associated to the radius is evaluated taking into account the uncertainties from different sources, as the extrapolation to the static point, the choice of the class of fitting functions and the range of the data sample. This error is shown to be even orders of magnitude larger than commonly assumed.
\end{abstract}

\maketitle

\section{Introduction}

The precise determination of the proton radius, where with proton radius we mean the square root of the mean square electric radius, $R_E\equiv\rr$, is object of intensive experimental and theoretical studies, especially since a recent experiment on Lamb shift in muonic hydrogen~\cite{Pohl:2010zza,Antognini:1900ns} measured a value which has large discrepancy with respect to previous measurements performed with electron atomic hydrogen as well as with electron scattering experiments (for a review, see Ref.~\cite{Bernauer:2014nqa}). The value measured using laser spectroscopy of muonic hydrogen, $R_E^{H^\mu}=0.84087(39)$ fm,
	has a relative error one order of magnitude lower than the previous datum. It is smaller, by more than seven standard deviations, with respect to the 2010-CODATA average value $R_E^{\rm CO}=0.8775(51)$ fm~\cite{Mohr:2015ccw}. Such a value has been obtained from a compilation of hydrogen atom spectroscopy data and electron-proton elastic scattering measurements at low momentum transfer. Latest experiments with electrons at Jlab~\cite{Zhan:2011ji} and MAMI~\cite{Bernauer:2010wm} confirm this last value. However, the situation is even more involved.  A recent measurement of hydrogen spectroscopy for the 2S-4P transition~\cite{Beyer:2017} gives the result $R_E^{H^e_{2S}}=0.8335(95)$ fm, compatible with muonic hydrogen, wheres an even more recent measurement of the 1S-3S transition~\cite{Fleurbaey:2017} still of atomic hydrogen concludes in the value $R_E^{H^e_{1S}} = 0.879(25)$ fm, again compatible with the CODATA result. 
\\
In this work we use only the data measured by the A1 Collaboration~\cite{Bernauer:2010wm}, that are the most precise available results at small momentum transfer. We critically revise the possibility to extract a precise information on the proton radius from elastic electron proton scattering.
\\
 The formalism describing the elastic scattering of electrons on protons is well known~\cite{Hofstadter:1956qs,Rosenbluth:1950yq}. In Born approximation, i.e., considering only one-photon exchange, the cross section is expressed in terms of two independent Lorenz scalar functions of $q^2=-Q^2<0$, where $q$ is the four-momentum of the intermediate photon. Such functions are called electromagnetic form factors (FFs);  the choice of these function is not unique. The most common choice, mainly because of their connection with charge and magnetic-moment spacial distributions of the proton, is  the electric and magnetic Sachs FFs parameterization, $G_E$ and $G_M$ (for  a recent review on the determination of FFs in space and time-like regions, see~\cite{Pacetti:2015iqa}). It is indeed the first $Q^2$-derivative of the electric FF at the limit $Q^2\to 0$, which is proportional to the proton radius. From the experimental point of view, the extraction of the proton radius requires an extrapolation to $Q^2=0$ of the measured observable, i.e., the elastic cross section, that, in the limit $q^2\to 0^-$, is dominated by the electric FF $G_E$. 
 Form factors are, in general, complex analytical functions of the single variable $q^2$, or equivalently $Q^2$, defined in the whole $q^2$ complex plane, with a cut, along the positive real axis (time-like region), from the pion threshold $q^2=(2M_\pi)^2$ up to $q^2=\infty$.\\
 The differential cross section for the process
 \be 
 e^- + p\to e^- +p\,,
\nen
 that is the electron-proton elastic scattering, in Born approximation and in the proton rest frame, where initial and final electron four-momenta are $k_{1,2}=\lt\omega_{1,2},\vec{k}_{1,2}\rt$, has the expression
 \be
 \frac{d\sigma}{d\Omega}
 =
\lt  \frac{d\sigma}{d\Omega}\rt_{\rm Mott}
\frac{\e(\tau,\theta_e)\, G^2_E\lt q^2\rt+\tau\, G^2_M\lt q^2\rt}{\e(\tau,\theta_e)\lt 1+\tau\rt}\,,
\label{eq:scatt-xs}
\en
where the first factor is the Mott point-like cross section
\be
\lt  \frac{d\sigma}{d\Omega}\rt_{\rm Mott}
=\frac{4\alpha^2}{\lt -q^2\rt^2}\,\frac{\omega_2^3}{\omega_1}
\lt 1-\frac{\vec{k}_1^2}{\omega_1^2}\sin^2\lt\frac{\theta_e}{2}\rt\rt
\simeq 
\frac{4\alpha^2}{\lt -q^2\rt^2}\,\frac{\omega_2^3}{\omega_1}
\cos^2\lt\frac{\theta_e}{2}\rt
\,;
\label{eq:mott}
\en
$\tau=-q^2/\lt 4M^2\rt>0$ is the normalized squared four-momentum;
$\theta_e$ is the electron scattering angle, and 
$\e(\tau,\theta_e)$ is the virtual photon polarization
\be
\e(\tau,\theta_e)=\frac{1}{1+2(1+\tau)\tan^2(\theta_e/2)}\,.
\label{eq:epsilon}
\en
The approximate identity in Eq.~\eqref{eq:mott} holds when
the mass, $M_e$, of the electron is negligible with respect to its initial energy $\omega_1$, and hence $\vec{k}_1^2=\omega_{1}^2-M_e^2\simeq\omega_1^2$. Moreover, in such an approximation, the four-momentum squared reads $q^2\simeq - 4\omega_1\omega_2\sin^2\lt\theta_e/2\rt$.
\\
The Born elastic scattering cross section of Eq.~\eqref{eq:scatt-xs} diverges as $1/\lt Q^2\rt^2$ when $Q^2\to 0^+$. This limit is fulfilled when, either the scattering angle, $\theta_e$, or the electron initial energy, $\omega_1$, vanishes, indeed: $Q^2=o\lt\theta_e^2\rt$ as $\theta_e\to 0$, and $Q^2=o\lt\omega_1^2\rt$ as $\omega_1\to 0^+$ (the vanishing of $\omega_1$ implies also $\omega_2\to 0^+$, since $\omega_2=O(\omega_1)$ as $\omega_1\to 0^+$). The physical reasons can be easily understood: either the incident particle is not deflected ($\theta_e=0$), and it does not "see" the target; or, if the energy is too low, the Coulomb interaction induces a capture process, a non-perturbative QED phenomenon, hence the breaking of the Born approximation.
\\
 Therefore, the determination of the proton radius requires an extrapolation of the scattering process and of the related formalism in the kinematical region where the electron is captured and an atom is formed.  The description of the trapping process is highly model dependent, as it involves different types of corrections, most of which can not be calculated exactly. Note that, in general, photo and electro-production reactions are studied separately, i.e., one can not describe experimental observables measured with a real photon, as the static limit of electro-production models; numbers of amplitudes and the spin structure of the matrix elements are different.  
 \\
The low-$Q^2$ scattering is also affected by model dependent corrections. As discussed here, not all related systematic uncertainties have been taken into account and correctly reflected in the official errors on the radius. 
%
%
\subsection{Coulomb corrections}
\begin{figure}[h!]
\begin{center}
	\includegraphics[width=0.5\textwidth]{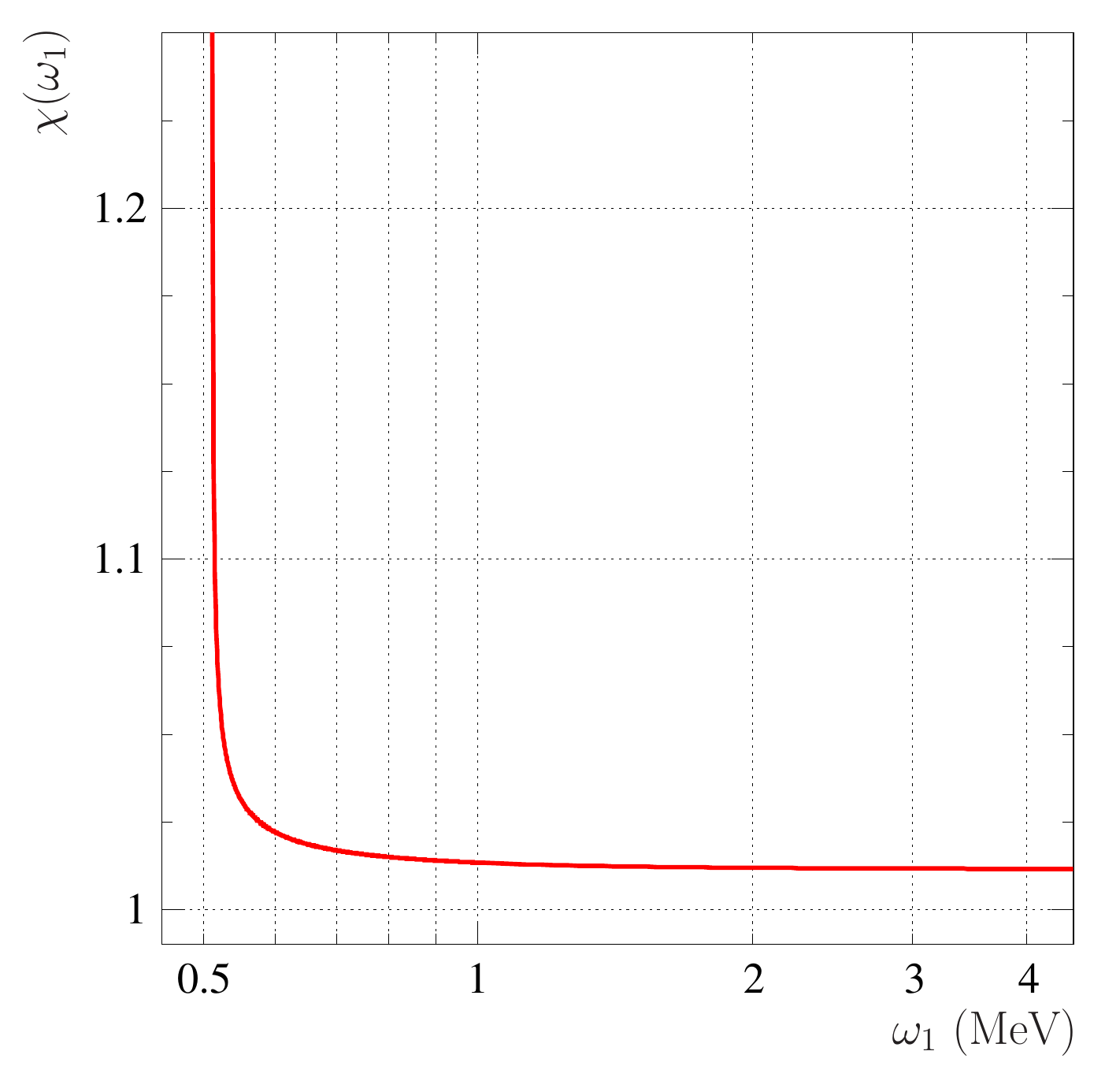}\vspace{-4mm}
\caption{\label{fig:coulomb}The Coulomb correction of the scattering cross section as a function of the initial electron energy.}
 \end{center}
\end{figure}

%
%
The importance of Coulomb corrections has been pointed out in Ref.~\cite{Lee:2014uia,Krachkov:2015uva}. Corrections due to multi-photon exchange should be taken into account at energies below 200 MeV, as they are of the order of percent, which is the size of the errors of the present cross section measurements. Screening effects become huge at very low energies, where the Born cross section needs to be corrected. To overcome this issue, one generally introduces a minimum scattering angle, $\theta_{e,\rm min}$, which is related to the impact parameter, and defined as the maximum value between the classical (c) and the quantum (q) expressions~\cite{Jackson:1975}
\be
\theta_{e,\rm min}^{\rm (c)}=\frac{2 e^2}{|\vec{k}_1| b\, \beta}\,,\hh\hh
\theta_{e,\rm min}^{\rm (q)}=\frac{\hbar}{|\vec{k}_1| b}\,,
\label{eq:thmin}
\en
where $b$ is the impact parameter, $\vec{k}_1$ and $\omega_1$, as defined above, the electron three-momentum and energy in the proton rest frame, and $\beta=|\vec{k}_1|/\omega_1$ the relative velocity. The Coulomb correction, that factorizes in the cross section expression, is given by~\cite{sommerfeld1921atombau,Sakharov:1948yq}
\begin{equation}
\chi(\omega_1)\equiv 
\left|\psi(0)\right|^2=\displaystyle\frac{\pi\alpha/\beta}{1-e^{-\pi\alpha/\beta}}\,,
\label{eq:chi}
\end{equation}
where the wave function $\psi(r)$ is the solution of the Schr\"odinger equation with a Coulomb potential, and hence it describes a non-relativistic system of two interacting charged particles.  
Such a correction represents an attractive (repulsive) interaction for opposite (same) sign charges and increases (decreases) abruptly the cross section up to $\omega_1=4$ MeV, which is, however, a value much lower than the energy of the considered experiments. Fig.~\ref{fig:coulomb} shows the behavior of the Coulomb correction as a function of the incoming electron energy $\omega_1$. The divergence in the low-energy region is due to the presence of a pole and hence a vertical asymptote at $\omega_1=M_e$, where the relative velocity $\beta$ vanishes. 
The cutoff angle, when the scattering does not occur anymore is $\theta_{e,\rm min} \sim 10^{-5}$ rad, corresponding to the four-momentum squared $Q^2\sim 10^{-5}$ MeV$^2$. Below such values the scattering formalism, even corrected, is not a relevant starting point to model the process.
%
%
 \subsection{Effect of the lepton mass}
\begin{figure}
\begin{center}
\includegraphics[width=0.49\textwidth]{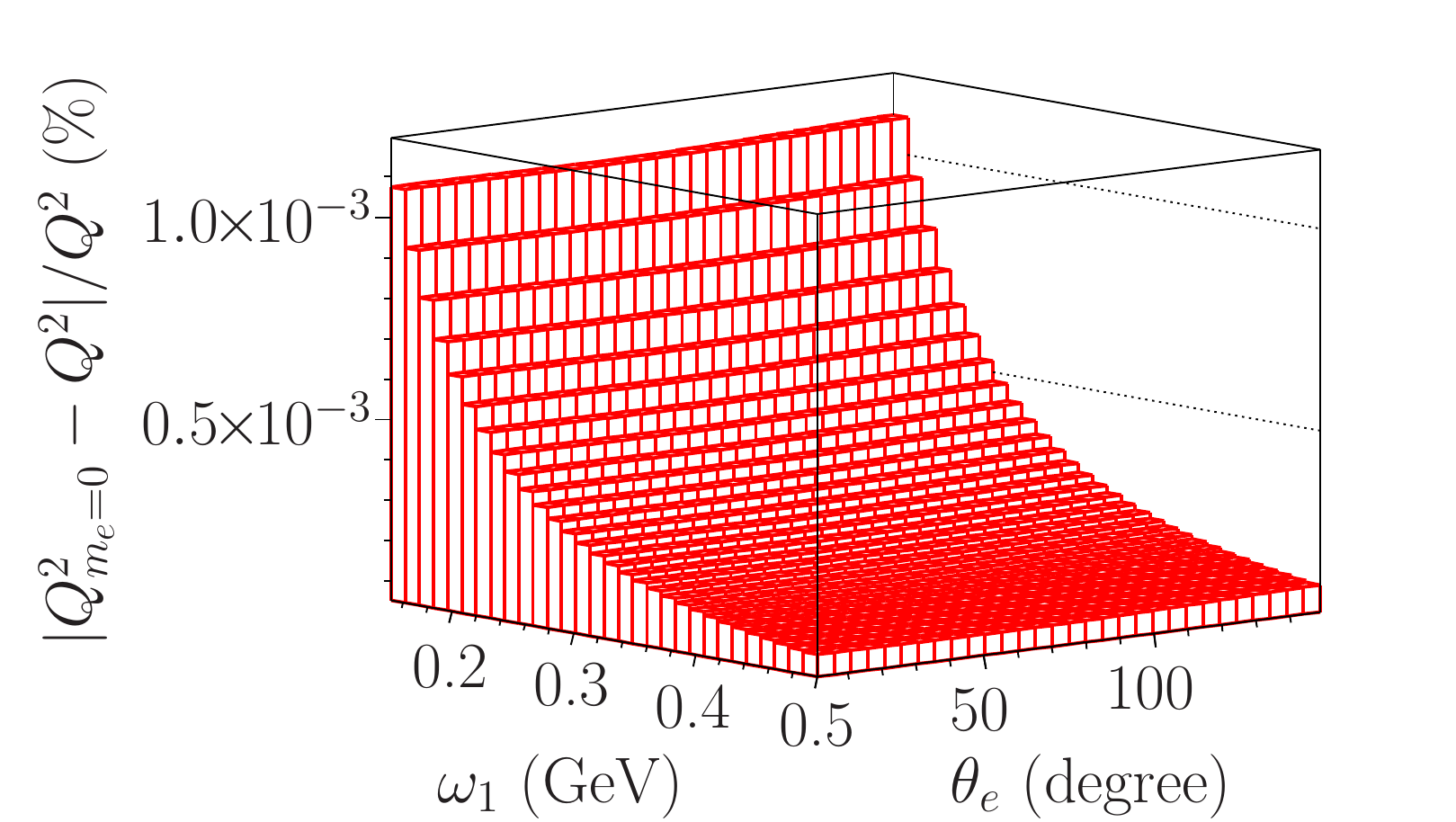}
\includegraphics[width=0.49\textwidth]{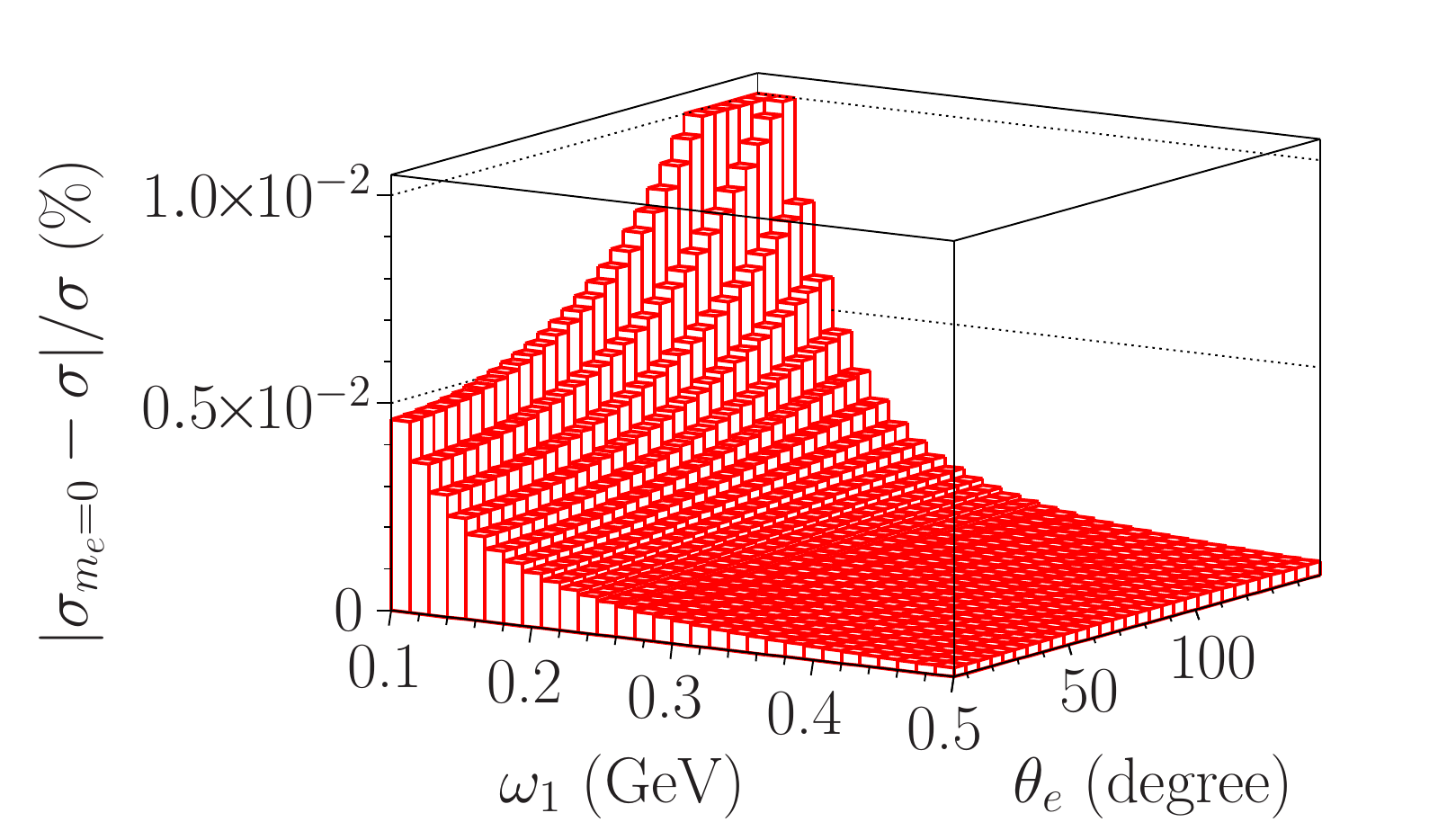}
\caption{\label{Rosenbluth1}Relative difference of $Q^2$ (left)  and $\sigma$ (right) taking and not taking into account the electron mass.}
	\end{center}
\end{figure}

The effect of the lepton mass in the calculation of lepton-proton cross section becomes important in the non relativistic limit, i.e., when the mass itself can not be neglected. Such an eventuality is verified at the MeV scale for muon-proton elastic scattering, as well as for inverse kinematics, where a proton beam collides with atomic electrons. The presence of mass terms modifies the kinematical variables and induces additional factors in the cross section. Fig.~\ref{Rosenbluth1} shows the relative differences on $Q^2$ and on the cross section, calculated with and without setting the electron mass to zero ($Q^2_{m_e=0}$, $\sigma_{m_e=0}$ and $Q^2$, $\sigma$). In the relevant kinematical range $0.1\le\omega_1\le 0.5$, the relative effects are $<10^{-5}$ and $<10^{-4}$ for the four-momentum squared and the cross section respectively. Note that, even though these are small effects for electron probes at the energies of interest, they are of the same order of the accuracy on the cross section measurements, $\sim 10^{-4}$, that is needed to extract values of the radius precisely enough to discriminate between the two reference values $R_E=0.84$ fm and $R_E=0.87$ fm.
\\
%
%
\section{Extracting the proton radius from $G_E\lt Q^2\rt$}
\label{subsec:radius}
In order to quantify the limitations in using scattering observables to extract the proton radius from low-$Q^2$ $G_E$ behaviour, we consider only the data from the experiment of Ref.~\cite{Bernauer:2010wm}. Indeed it gives until now the most precise data in the largest low-$Q^2$ range. Moreover, considering only one set of data allows to avoid normalization issues among different experiments.
\\
In the original paper~\cite{Bernauer:2010wm} two kinds of analyses were performed. The first method is based on a global fit of the cross section, assuming pre-defined functional forms for FFs. Several classes of functions have been used, giving consistent results on the radius as shown in Fig.~10.1 of Ref.~\cite{Bernauer:2010zga}, where the first derivatives of different parameterizations normalized to the dipole derivative, are represented as function of $Q^2$. The fact that different functional forms for FFs have similar first derivative and hence give similar radii is not a measure of the stability of the result, but rather of its model dependence. Indeed, as it is argued here, the choice of predefined functions puts serious constraints on the determination of the radius, as can be inferred by considering, instead, FFs extracted from the differential cross section by using the standard Rosenbluth technique.
\\
This is, in fact, the second method proposed in Ref.~\cite{Bernauer:2010zga}. In particular the Rosenbluth analysis consists in extracting the FFs from the slope and the intercept of the reduced cross section, which is defined starting form the expression of Eq.~\eqref{eq:scatt-xs}, by isolating the part depending on the FFs, i.e., 
\be
\sigma_{\rm red}(\e,q^2)\equiv
\frac{d\sigma/d\Omega}{\lt d\sigma/d\Omega\rt_{\rm Mott}}\,\e\lt1+\tau\rt
=\e\,G_E^2\lt q^2\rt+\tau\,G_M^2\lt q^2\rt\,.
\label{eq:red-xs}
\en
At fixed $q^2$ the reduced cross section depends only on the virtual photon polarization $\epsilon$ given in Eq.~\eqref{eq:epsilon}. 
Although Rosenbluth data have larger errors and their range does not extend to the lowest measured $Q^2=0.0005$ GeV$^2$, they are in principle preferable, as they are not aligned along a pre-established function, which unavoidably induces a bias also on the derivative. By relying only on the one-photon exchange assumption, the Rosenbluth method can be considered a model-independent procedure to extract FF values from cross section data. In our analysis we do not consider the first low-$Q^2$ FFs values extracted with the Rosenbluth technique, because they have been arbitrarily rescaled, so that only data  at $Q^2~\ge~0.0202 $ GeV$^2$ have been taken into account. 
%
%
%
%
%

The weakness of the procedure of extracting the proton radius from FF data is proven by the fact that several re-analysis based on fits to the these data, in different $Q^2$ ranges and/or with different functional forms, give radii in a large range including the extreme values from hydrogen and muonic atoms, $R_E^{H^\mu}$ and $R_E^{H^e_{1S}}$~\cite{Kraus:2014qua,Lorenz:2014yda,Horbatsch:2016ilr,Griffioen:2015hta}. 
 In Ref.~\cite{Bernauer:2016ziz} a number of fits and results is presented and discussed in terms of statistical analysis. Different aspects are brought to the attention of the reader: the cutoff set for the low-energy fits, the degree of polynomial expansion, the poor description of the well known dipole function. At low $Q^2$ a development with polynomials on $G_E$, is compared to the Taylor expansion of the FF in terms of the radius. The test is done on "pseudo data" that have been built in order to discriminate the two hypotheses of a "small", $R_E=0.83$ fm, and a "large" radius, $R_E=0.87$ fm, using a given functional form for the FFs. By considering low-$Q^2$ points, the extrapolation at $Q^2=0$ leads to a large value of the radius, compatible with CODATA. However, by taking a more restricted range, that excludes low-$Q^2$ points, i.e., data with $Q^2\le 0.1$ GeV$^2$, one gets a smaller value of the radius, even smaller than that from muonic atoms, $R_E^{H^\mu}$. Any intermediate solution can be found by choosing the opportune $Q^2$ interval and fit function, e.g., by considering polynomials of different degree. 
 
 We present a procedure for extracting the proton radius from scattering data, which has two main features.
 \begin{itemize}
 \item The first derivative of the electric FF is obtained by fitting directly the data on the discrete increment of $G_E$, rather than as the first derivative of the fitting function.
 \item The fit functions are polynomials whose coefficients are constrained by a first-principle condition driven by a genuine charge density~\cite{Sick:2018fzn}.
 \end{itemize}
Among the many analyses that can be found in literature, the direct extraction of the derivative, i.e., of the discrete increment of the FF from the data has not been yet done. This is the object of the present paper, with particular attention to the errors attributed to this quantity.
%
\subsection{Sachs form factors}
The Sachs FF $G_C(q^2)$ ($C=E,M$ for the electric and magnetic FF respectively) is a Lorenz scalar function of $q^2$, the four-momentum squared of the photon. In the reference system, called Breit frame, where the scattering proceeds with no energy exchange, hence $q=(0,\vec q)$, the FF can be interpreted as the three-dimensional Fourier transform of the corresponding charge density $\rho_C(\vec r)$, i.e.,
\be
G_C(q^2)=\int d^3\vec{r}\,\rho_C(\vec r) \,e^{-i\vec{q}\cdot\vec{r}}\,.
\nen
By considering a spherical density, $\rho_C(\vec{r}) =\rho_C(r)$, depending only the norm $r=|\vec{r}|$, the angular and radial integrations factorize as
\be
G_C(q^2)=2\pi\int_0^\infty r^2 dr\,\rho_C(r) \int_{-1}^1 d\cos(\theta)\,e^{-i|\vec{q}|\,r\,\cos(\theta)}\,,
\label{eq:Fourier1}
\en
where $\theta$ is the angle between the position $\vec{r}$ and three-momentum $\vec{q}$. In the Breit frame: $Q^2=-q^2={\vec{q}}^{\,2}\ge 0$, that also implies $|\vec{q}|=Q$, hence we re-define the momentum dependence of FFs as: $G_C(q^2)\to G_C(Q^2)$.
\\
By making the $d\cos(\theta)$ integration of the Fourier transform of Eq.~\eqref{eq:Fourier1}, the FF reads
\be
\wg_C(Q^2)=\frac{4\pi}{Q}\int_0^\infty r\rho_C(r) \sin(Qr)\,dr\,.
\label{eq:fourier1}\en
At this stage, the uniformly convergent Taylor series of the sinus function can be exploited to obtain the series for $\wg_C\lt Q^2\rt $ in $Q^2=0$ as
\be
\wg_C\lt Q^2\rt\ug
4\pi \sum_{j=0}^\infty
(-1)^n\frac{\lt Q^{2}\rt^n}{(2n+1)!}
\int_0^\infty r^{2n+2}\rho_C(r)\,dr
=
G_C(0)
\sum_{n=0}^\infty
(-1)^n\frac{\mean{r^{2n}}_C}{(2n+1)!}\lt Q^{2}\rt^n
\!\!,
\label{eq:poly}
\en
where the $2n$-mean radius is defined as
\be
\mean{r^{2n}}_C\equiv \frac{\ds 4\pi\int_0^\infty r^{2n+2}\rho_C(r)\,dr}
{\ds 4\pi\int_0^\infty r^{2}\rho_C(r)\,dr}
\,,\hh n\in\N\,.
\nen
The normalization is to the total charge "$C$", which is given by taking the limit $Q\to 0^+$ of Eq.~\eqref{eq:fourier1}, i.e., by integrating the density over all space, and it represents the value of the FF at $Q^2=0$, 
\be
G_C(0)=4\pi\int_0^\infty r^{2}\rho_C(r)\,dr\,,
\nen
and hence the $2n$-mean radius is also defined as
\be
\mean{r^{2n}}_C
=
\frac{1}{G_C(0)}4\pi\int_0^\infty r^{2n+2}\rho_C(r)\,dr
\,.
\nen
Moreover, by following the Cauchy-Hadamard theorem, the convergence radius in $Q^2$ is
\be
Q^2_C=\left\{\limsup_{n\to\infty}\lq\frac{\mean{r^{2n}}_C}{(2n+1)!}\rq^{1/n}\right\}^{-1}\,.
\label{eq:CH}
\en 
Since FF are analytic functions of $Q^2$, their power series expansions are unique and hence the one given in Eq.~\eqref{eq:poly} does coincide with the Taylor series of $\wg_C(Q^2)$, centred at $Q^2=0$,
\be
\wg(Q^2)=\sum_{n=0}^\infty \frac{1}{n!}\left.\frac{d^n\wg_C}{d\lt Q^2\rt^n}\right|_{Q^2=0}\,\lt Q^2\rt^n\,.
\nen
The coincidence of the two series means that homologous coefficients are equal to each other, so that mean values of $r^{2n}$ are proportional to the $n$-th derivatives of the FF, i.e,  
\be
\left\langle r^{2n}\right\rangle_C
=(-1)^n\frac{(2n+1)!}{n!}
\frac{1}{G_C(0)}
\left.\frac{d^n\wg_C}{d\lt Q^2\rt^n}\right|_{Q^2=0}\,,\hh
n\in\N\,.
\nen
In particular, with $n=1$ we obtain, assuming sphericity of charge density, the expression of the mean square radius in terms of the normalized derivative of the electric FF at $Q^2=0$, or equivalently of its logarithmic derivative,
\be
\left\langle r^2\right\rangle_C
=-\frac{6}{G_C(0)}\left.\frac{d\wg_C}{dQ^2}\right|_{Q^2=0}
=-6\left.\frac{\ln[\wg(Q^2)]}{dQ^2}\right|_{Q^2=0}
\,.\label{eq:r-der}
\en
This represents the standard relation exploited to extract the proton radius from elastic scattering data.
\\ 
As an example, we consider the well known case of a FF behaving as the standard dipole formula
\be
G_D(Q^2)=\lt 1+\frac{Q^2}{0.71\,{\rm GeV}^2} \rt^{-2}\,.
\label{eq:dip}
\en
The power series ($G_D(0)=1$) is
\be
G_D(Q^2)=\sum_{n=0}^\infty (-1)^n \frac{n+1}{\lt 0.71\,\,{\rm GeV}^2\rt^n} \, \lt Q^2\rt^n\,,
\label{eq:dipole-series}
\en
so that the mean radii, with $n\in\N$, are
\be
\mean{r^{2n}}_D=\frac{(n+1)(2n+1)!}{\lt 0.71\,\,{\rm GeV}^2\rt^n}
=
\frac{(n+1)(2n+1)!}{12^n} R_D^{2n}\,,
\label{eq:raggi}
\en
where $R_D$ is the dipole radius, obtained by taking the square root of the first identity with $n=1$, i.e.,
\be
R_{D} =
\sqrt{\frac{12}{ 0.71\,\,{\rm GeV}^2}}\simeq
4.111\,{\rm GeV}^{-1}
\simeq 0.811\,{\rm fm}\,.
\nen
The conversion GeV$^{-1}\leftrightarrow$ fm, here and elsewhere, is performed by assuming natural units, and hence by means of the identity $\hbar c\equiv\lt 0.1973269788\pm 12\rt \,{\rm GeV}\,{\rm fm}=1$~\cite{pdg}, which implies
\be
1\,{\rm GeV}^{-1}=\lt 0.1973269788\pm 12\rt \,{\rm fm}\,.
 \label{eq:units}
 \en
 The convergence radius of the series given in Eq.~(\ref{eq:dipole-series}) corresponds to the distance from the center of the series, $Q^2=0$, and the unique pole of the dipole which lies at $Q^2=-0.71$ GeV$^2$. Indeed, from Eq.~\eqref{eq:CH},
\be
Q^2_{C}=
\left\{\limsup_{n\to\infty}\lq\frac{n+1}{\lt0.71\,{\rm GeV}^2\rt^n}\rq^{1/n}\right\}^{-1}
=0.71\,{\rm GeV}^2\,.
\nen
%
%
%
\subsection{Polynomial description of $\wg_E(Q^2)$}
%
Polynomial expansions for the normalized electric FF and its first derivative can be obtained from Eq.~(\ref{eq:poly}) and are 
\be
\frac{\wg_E\lt Q^2\rt}{G_E\lt 0\rt}
\ug
1-\frac{1}{3!}R^2_E Q^2 +
\sum_{n=2}^\infty (-1)^n\frac{\langle r^{2n}\rangle_E}{(2n+1)!}\lt Q^2\rt^{n}\,,
\no\\
\frac{1}{G_E\lt 0\rt}
\frac{d\wg_E}{dQ^2}\lt Q^2\rt\ug
-\frac{1}{3!}R^{2}_E  +
\sum_{n=2}^\infty (-1)^n\frac{\langle r^{2n}\rangle_E}{(2n+1)!} \,n\lt Q^2\rt^{n-1}
\,.
\nen
The coefficients, which are proportional to the mean values of even powers of the radius, can be constrained using the generalized Schwarz's inequality of Eq.~\eqref{eq:general}, obtained in App.~\ref{app:Schwarz},
\be
\langle r^{2n}\rangle_E\ge \langle r^{2}\rangle^{n}_E = R_E^{2n}\,.
\nen
Such a relation is particularly useful and can be imposed in a fitting procedure by defining the set of non-negative parameters $\left\{R^2_E,\Delta_2^2,\Delta_3^2,\dots,\Delta_k^2,\dots\right\}\subset [0,\infty)$, with the dimension of length squared, through the relation 
\be
\la r^{2n}\ra_E \equiv \lt R_E^{2}+\Delta_n^2\rt^n\,.
\nen 
It follows that the polynomial descriptions of the FF and its first derivative, depending on the $N+1$ parameters of the set $\left\{\Delta_0^2,R^2_E,\Delta_2^2,\Delta_3^2,\dots,\Delta_N^2\right\}$ are then obtained by truncating the series at a given order $N\ge 2$, as
\be
\wg_E\lt Q^2;\Delta_0^2,R_E^2,\Delta_2^2,\Delta_3^2,\dots,\Delta_N^2\rt
\ug\lq
1-\frac{1}{3!}R_E^2 Q^2 +
\sum_{n=2}^N (-1)^n\frac{\lt R_E^2+\Delta^2_n\rt^n}{(2n+1)!}\lt Q^2\rt^{n}
\rq \Delta_0^2
\,,\no\\
&&\label{eq:fit-fun}\\
\frac{d\wg_E}{dQ^2}\lt Q^2;R_E^2,\Delta_2^2,\Delta_3^2,\dots,\Delta_N^2\rt\ug
\lq
-\frac{1}{3!}R_E^2  +
\sum_{n=2}^N (-1)^n\frac{\lt R_E^2+\Delta^2_n\rt^n}{(2n+1)!} n\lt Q^2\rt^{n-1}\rq\Delta_0^2
\,.
\nen
The dimensionless parameter $\Delta_0^2$, representing the FF value at $Q^2=0$, has been introduced as a normalization factor to correct eventual experimental normalization issues. Indeed, from the theoretical point of view, as a consequence of the total charge conservation, the electric FF at $Q^2=0$ in units of the positron charge $|e|$ is equal to one.\\ 
The polynomial describing the derivative, given in the second expression of Eq.~\eqref{eq:fit-fun}, depends only on $N$ free parameters, since the overall factor $\Delta_0^2$, that could not be determined independently from all the other parameters, is fixed to the value determined by fitting FF data with the first polynomial of Eq.~\eqref{eq:fit-fun}.
%

\section{Precision on the differential cross section}
\begin{figure}[h!]
\begin{center}
	\includegraphics[width=0.48\textwidth]{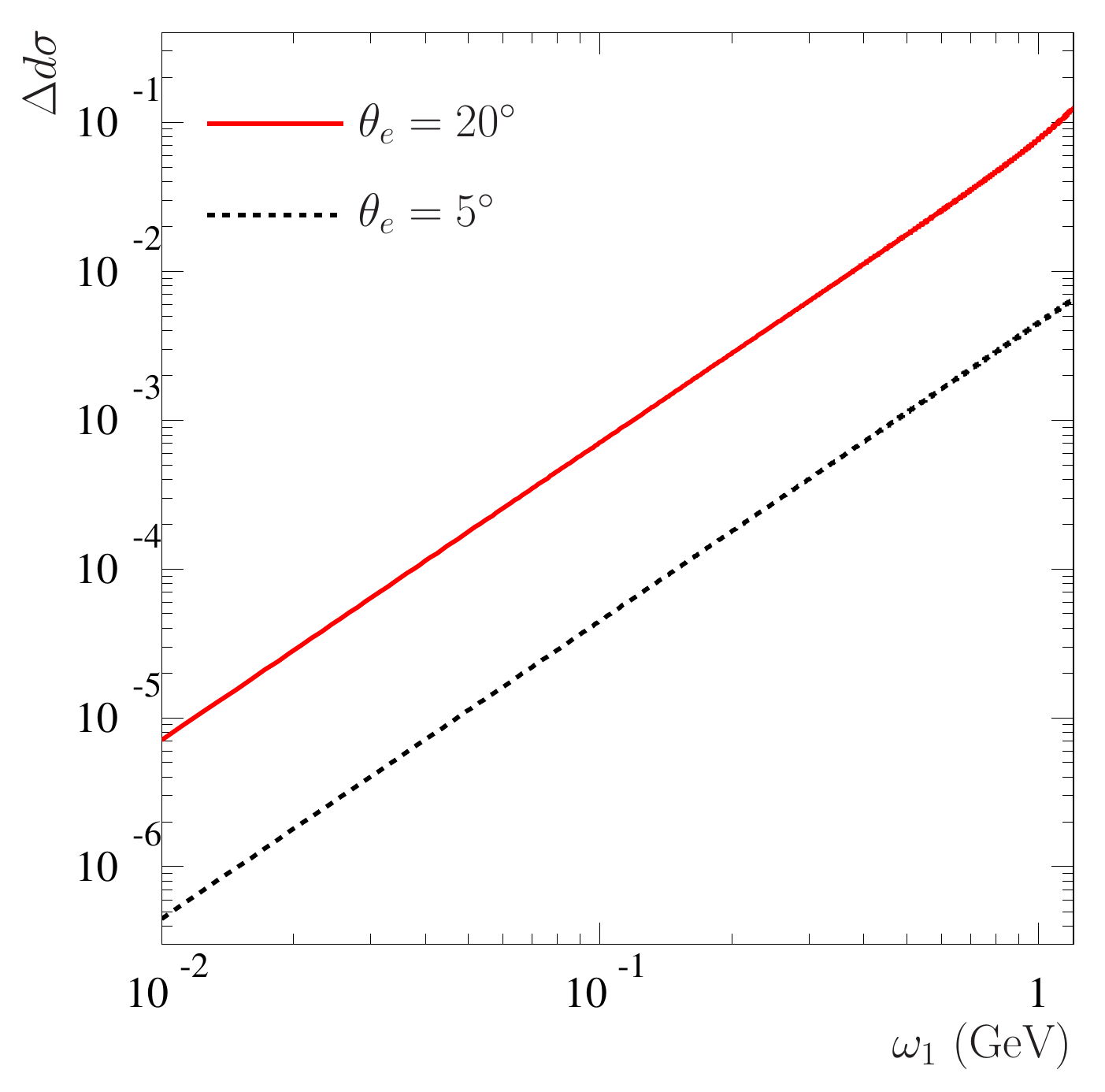}\hfill%
\includegraphics[width=0.48\textwidth]{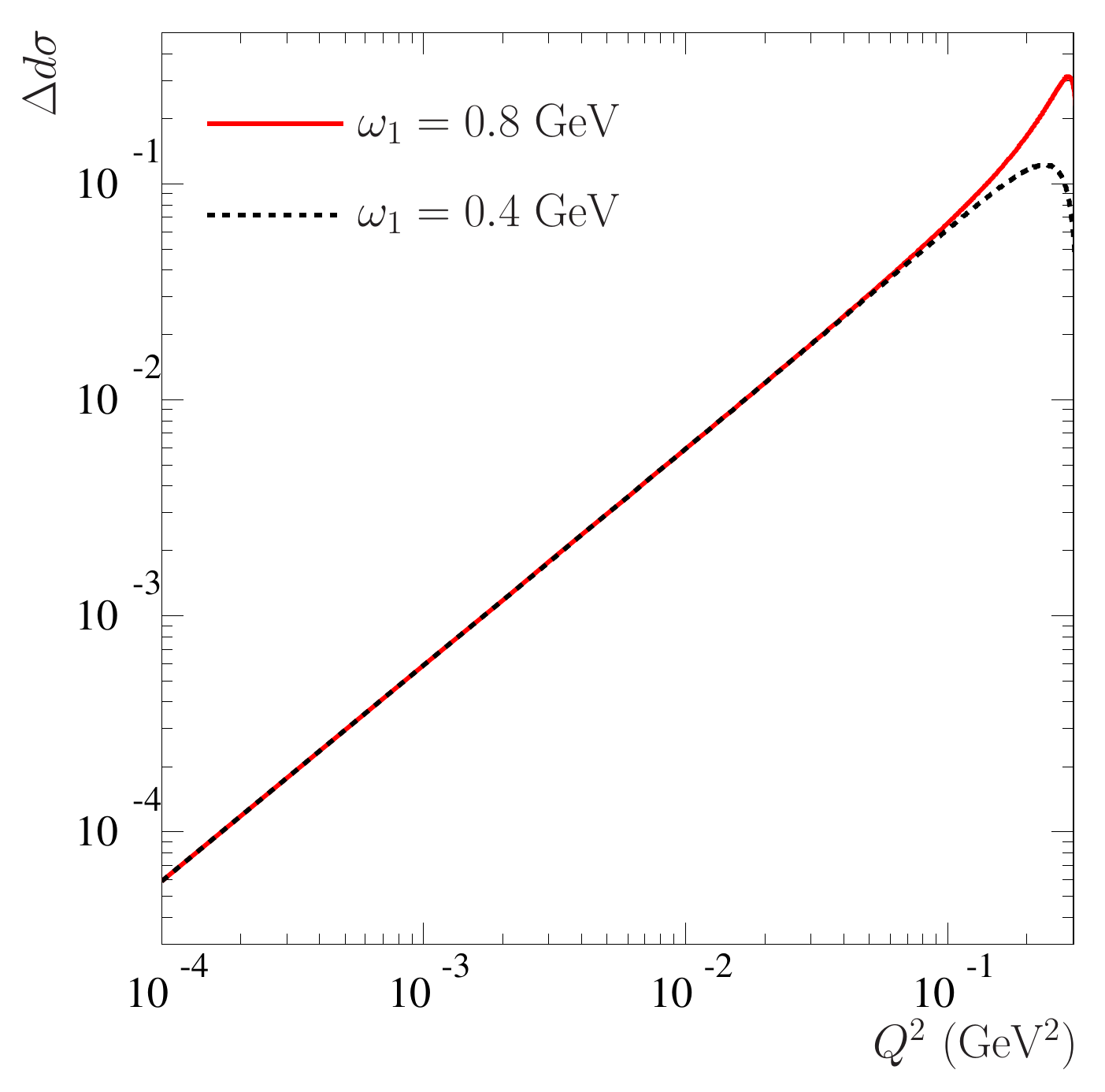}
\caption{Relative difference of the elastic cross section, for the two values of the proton radius, -as a function of the incident energy $\omega_1$, left panel, for two electron scattering angles:  $\theta_e=5^\circ$ (black dashed black line), $\theta_e=20^\circ$ (solid red line); - as a function of $Q^2$, right panel, at two values of the incident energy $\omega_1=0.4$ GeV (dashed black line) and $\omega_1=0.8$ GeV (solid red line).}
\label{fig:radius}
\end{center}
\end{figure}

%
The Rosenbluth technique~\cite{Rosenbluth:1950yq} consists in extracting FF values by fitting the data on the reduced cross section at fixed $Q^2$ as a function of the virtual photon polarization \e, Eq.~\eqref{eq:epsilon}, or equivalently of the scattering angle $\theta_e$. By using the expression of $\sigma_{\rm red}(\e,Q^2)$ given, in Born approximation, in Eq.~\eqref{eq:red-xs}, $G_E^2(Q^2)$ and $G_M^2(Q^2)$ have been obtained as the slope and the intercept (multiplied by $\tau$) of the behavior, linear in \e, of the reduced cross section.
\\
It has been assumed for a long time that the proton electric FF, as well as the magnetic FFs of proton and neutron, normalized to their magnetic moment, had the $Q^2$-dipole dependence of Eq.~\eqref{eq:dip}, whereas the neutron electric FF was essentially zero. 
\\
To quantify the level of precision that is needed to be achieved on the elastic cross section measurements, in order to discriminate among the radii measurements, we start from the expression of the differential cross section, Eq.~\eqref{eq:scatt-xs}, and then we consider the two extreme values of the radius
\begin{enumerate}
\item $R_{E(a)} =0.84$ fm, consistent with atomic measurements (except Ref.~\cite{Fleurbaey:2017});
\item $R_{E(b)}=0.88$ fm, similar to the one recommended by the 2010-CODATA review~\cite{Mohr:2015ccw}. 
\end{enumerate}
We use the following linear parametrization of FFs
\be
\label{eq:eqFF}
G_E^{a,b}(Q^2)= 1-  \frac {R_{E(a,b)}^2}{6(\hbar c)^2}Q^2\,,\hh 
G_M^D(Q^2)=\mu_p\lt  1- 2\frac{Q^2}{0.71\,{\rm GeV}^2}\rt\,,
\label{eq:FFsab}\en
where $\mu_p$ is the proton magnetic moment; i.e., we consider two possible expressions for the electric FF, corresponding to the radii $R_{E(a)}$ and $R_{E(b)}$, and a single expression for the magnetic FF. Using the expression of Eq.~\eqref{eq:scatt-xs}, we calculate the difference between the differential cross sections corresponding to the two different values of the proton radius, $R_{E(a)}$ and $R_{E(b)}$, as
\be
\Delta d\sigma \ug
\frac{d\sigma^ a/d\Omega - d\sigma^b/d\Omega}
{d\sigma^a/d\Omega }  
=
\frac{\sigma^{a}_{\rm red}\lt\e,Q^2\rt-\sigma^{b}_{\rm red}\lt\e,Q^2\rt}{\sigma^{a}_{\rm red}\lt\e,Q^2\rt}
=
\frac{{G^a_E}^2(Q^2) -{G_E^b}^2(Q^2)}{{G_E^a}^2(Q^2)+\tau {G_M^D}^2(Q^2)/\e}
\,,
\nen
where $d\sigma^{a,b}/d\Omega$ represents the differential cross section with the FFs $G_E^{a,b}$ and $G_M^{D}$ of Eq.~\eqref{eq:FFsab}. The results are shown in Fig.~\ref{fig:radius}, as function of energy of the incoming electron, $\omega_1$, (left panel) for two values of the electron scattering angle $\theta_e=5^o,\,20^o$ and of $Q^2$ (right panel), for two values of the incoming electron energy $\omega_1=0.4,\,0.8$ GeV. We use the kinematical relation among $Q^2$, $\omega_1$ and $\sin^2(\theta_e/2)$
\be
Q^2=4\omega_1\lt\omega_1-\frac{Q^2}{2M_p}\rt\sin^2(\theta_e/2)\,,
\nen 
where $M_p$ is the proton mass. Such a relation allows to determine one of three quantities once the other two are fixed. One can see that, in the approximation of Eq.~\eqref{eq:eqFF}, in order to discriminate the two extreme values of the radius, for $Q^2=10^{-3}$ GeV$^2$, measurements at the sub per mille precision are needed, as the relative difference, $\Delta d\sigma$, is lower than $10^{-3}$.
\\
The cross section dependence on the radius, i.e., on the first derivative of the electric FF, does increase as the incident energy, and/or the scattering angle and/or the squared four-momentum $Q^2$ increase. As FFs depend only on $Q^2$ in the one-photon-exchange approximation, the cross section at low  $Q^2$ is practically insensitive to the energy. These results show that the required precision on a cross section measurement to discriminate the two values of the proton radius is much larger than the percent, that is, for example, the precision inherent to  the calculations of radiative corrections at first order. Moreover, the possibility of obtaining precise values of the radius is even more weakened by the fact that such values do not rely on the cross section itself, but rather on the limit of the derivative of the FF for $Q^2\to 0^+$.

The parametrizations of Eq.~\eqref{eq:eqFF} are based on truncated expansions, where the neglected terms become important at large $Q^2$. Therefore we face such physical problem: the extrapolation of the measured cross section for  $Q^2\to 0^+$, requires low-$Q^2$ measurements with extremely challenging precision, whereas the measurements at large $Q^2$, where the sensitivity to FFs is higher, are affected by a larger uncertainty than in the low-$Q^2$ extrapolation.
%
%
%
 \section{The data}
\label{sec:data}
\begin{figure}[h!]
\begin{center}
	\includegraphics[width=0.48\textwidth]{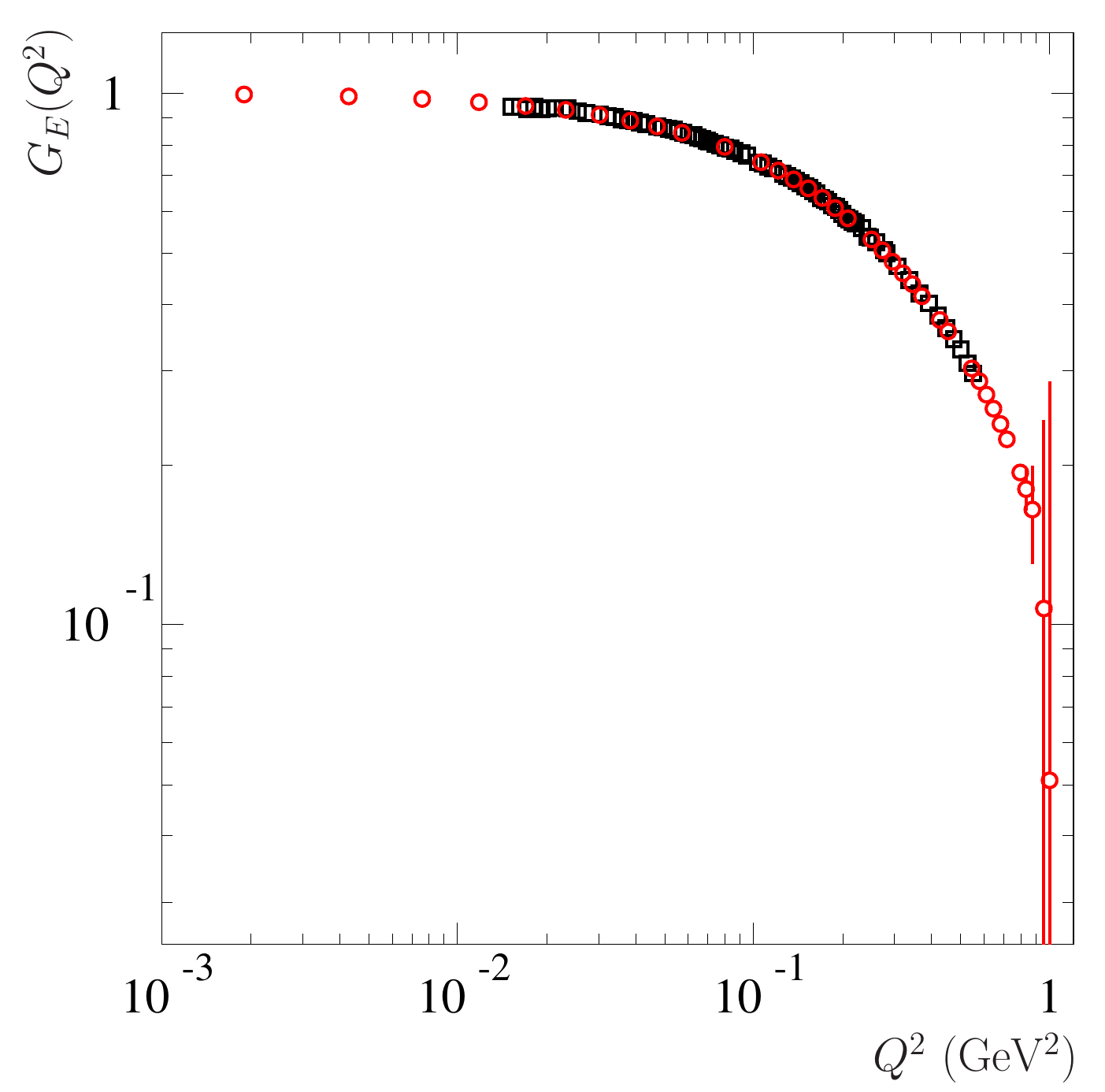}\hfill%
\includegraphics[width=0.48\textwidth]{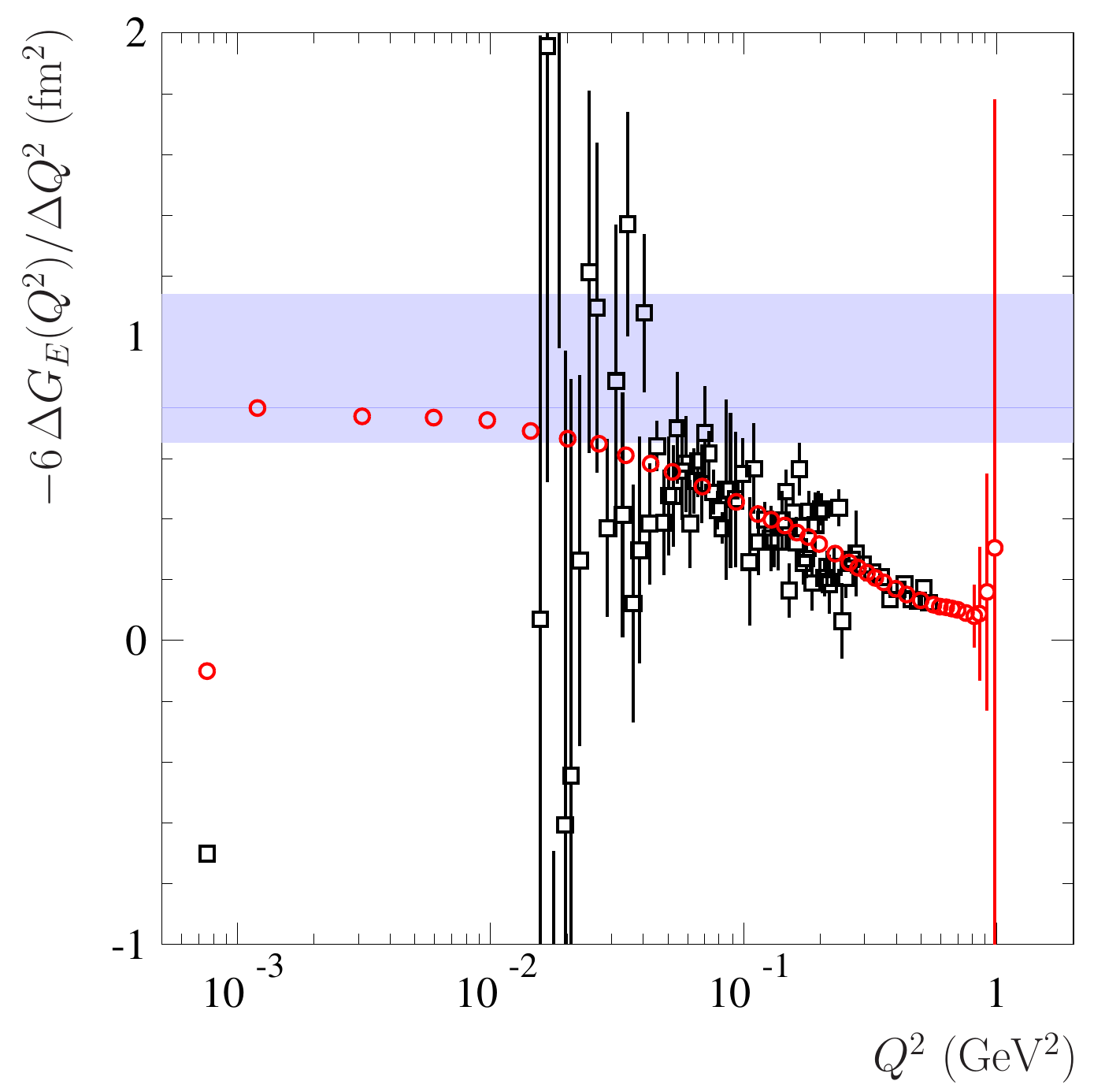}
	\caption{\label{fig:data-ge-dge} Left panel: Spline, red circles, and Rosenbluth data, black squares, on the electric proton FF. Right panel: Spline, red circles, and Rosenbluth data, black squares, on the discrete derivative of the electric proton FF, multiplied by $(-6)$ to obtain the quantity directly related to the mean square radius. Data are from Ref.~\cite{Bernauer:2010wm}. The light-blue and blue areas indicate radius intervals $0.80\le R_E\le 0.92$ fm and $0.84\le R_E\le 0.88$ fm, respectively.}
\end{center}	
\end{figure}

\begin{figure}[h!]
\begin{center}
	\includegraphics[width=0.48\textwidth]{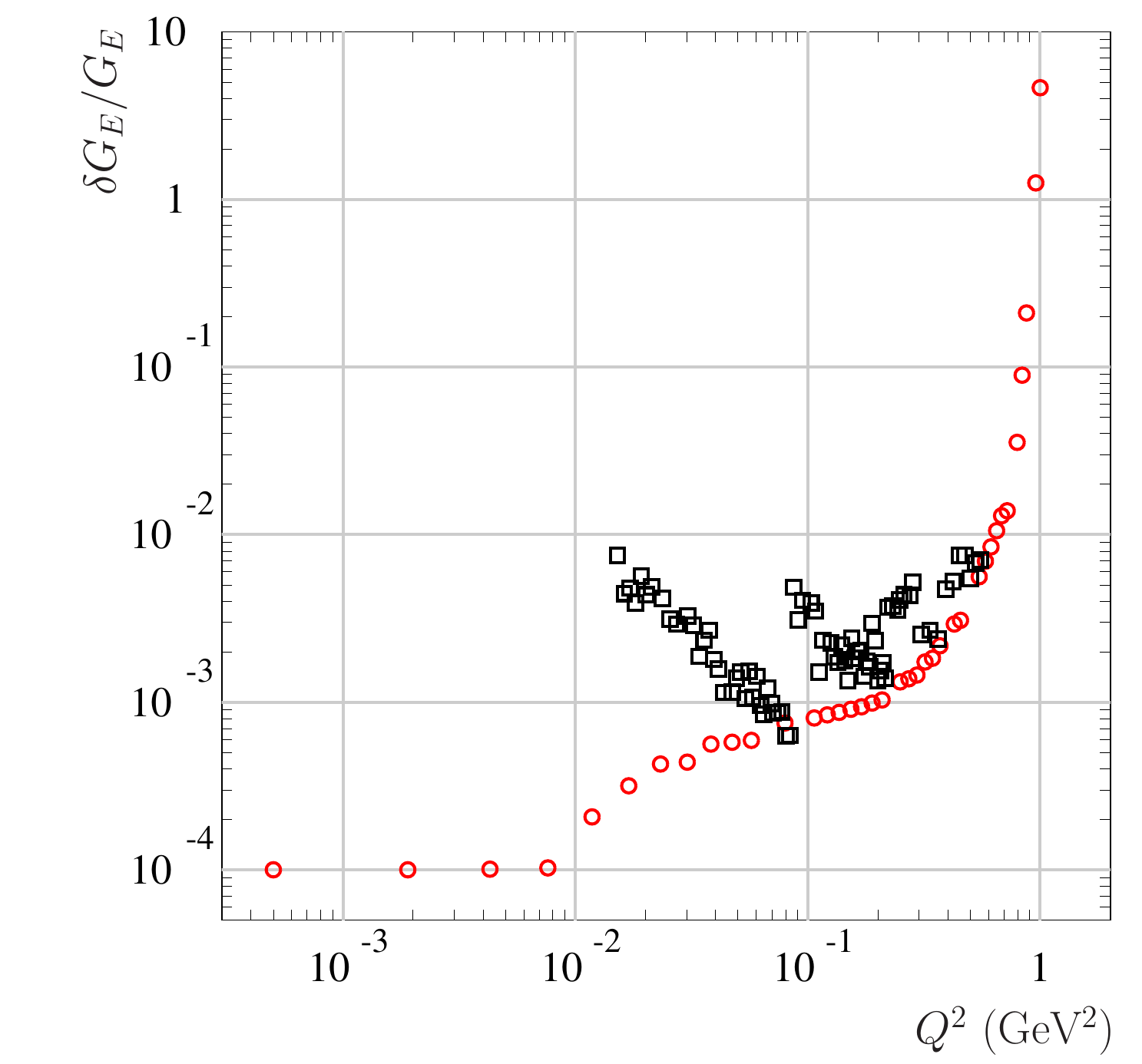}
	\hfill%
	\includegraphics[width=0.48\textwidth]{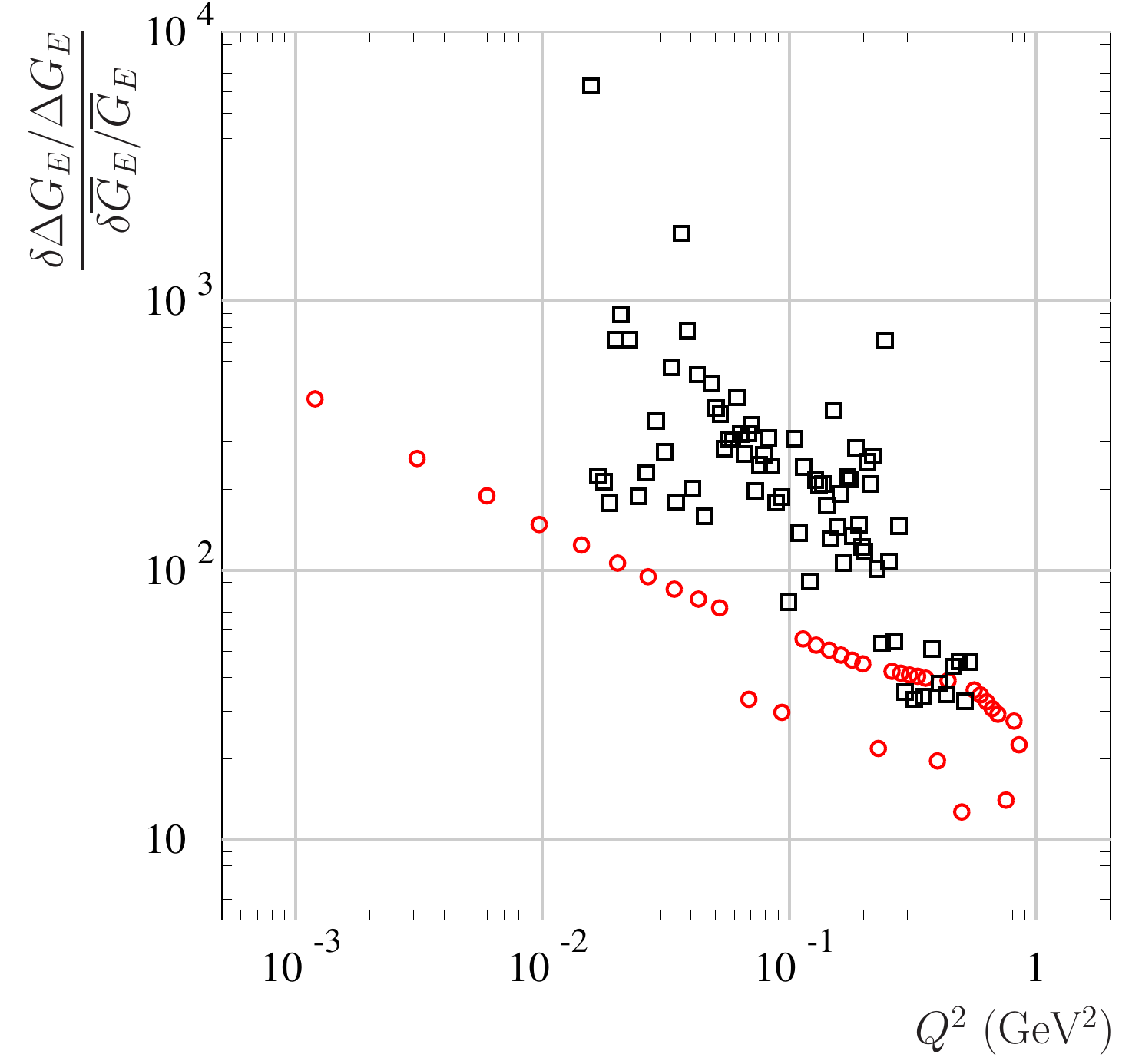}\caption{\label{fig:delta-delta}Ratio of relative errors on the discrete derivative and $G_E$, for spline (red circles) and Rosenbluth (black squares) data.}
\end{center}
\end{figure}

%
As discussed in Sec.~\ref{subsec:radius}, our study has been performed by using data from the only A1 experiment~\cite{Bernauer:2010wm}. In particular, the A1 Collaboration extracted FF values from differential cross section data by means of two procedures: the first is based on a global fit to the cross section, assuming a pre-defined analytic functional form for FFs; the second consists in the Rosenbluth technique, described in Sec.~\ref{subsec:radius}. In the following the two sets of data, obtained by means of these procedures, have been called "spline data" and "Rosenbluth data". 
\\
The two sets of data on $G_E\lt Q^2\rt$ are shown in the left panel of Fig.~\ref{fig:data-ge-dge}, red circles and black squares represent spline and Rosenbluth data respectively. There are two main differences between these data sets:
\begin{itemize}
\item the $Q^2$ range, $\lq 0.0005\,{\rm GeV}^2,1.0\,{\rm GeV}^2\rq$ for spline and $\lq 0.0152\,{\rm GeV}^2,0.5524\,{\rm GeV}^2\rq$ for Rosenbluth data;
\item Rosenbluth data have larger relative errors as can be seen in the left panel of Fig.~\ref{fig:delta-delta}.
\end{itemize}
Data on the discrete derivative of the FF are obtained from the original sets $\left\{Q_j^{2S},G^S_{E,j},\delta G^S_{E,j}\right\}_{j=1}^{N_S}$ and $\left\{Q_j^{2R},G^R_{E,j},\delta G^R_{E,j}\right\}_{j=1}^{N_R}$ of $N_S=38$ spline, apex "$S$", and $N_R=77$ Rosenbluth data points, apex "$R$", respectively, as the finite ratios and mean $Q^2$ values
\be
\Delta G_{E,j}^{S,R}=\frac{G^{S,R}_{E,j+1}-G^{S,R}_{E,j}}{Q^{2S,R}_{j+1}-Q^{2S,R}_{j}}\,,\hh
\delta \Delta G^{S,R}_{E,j}=
\frac{\sqrt{\lt\delta G^{S,R}_{E,j+1}\rt^2+\lt\delta G^{S,R}_{E,j}\rt^2}}{Q^{2S,R}_{j+1}-Q^{2S,R}_{j}}\,, \hh
 \ov{Q}^{2S,R}_j=\frac{Q^{2S,R}_{j+1}+Q^{2S,R}_{j}}{2}\,,
\label{eq:discreto}
\nen
with $j=1,\dots,N_{S,R}-1$, hence $\left\{\ov{Q}_j^{2S},\Delta G^S_{E,j},\delta\Delta G^S_{E,j}\right\}_{j=1}^{N_S-1}$, $\left\{\ov{Q}_j^{2R},\Delta G^R_{E,j},\delta\Delta G^R_{E,j}\right\}_{j=1}^{N_R-1}$ represent the two sets of discrete derivatives.
{\color{red}We have considered only the first-order central finite difference for two main reasons: it entails a lower error,  depending only on two measured values of the FF; the use of higher-order estimators is prevented by non-homogeneous distribution of the experimental squared four-momenta $Q^{2S,R}_j$.}

 The right panel of Fig.~\ref{fig:data-ge-dge} shows the spline (red circles) and Rosenbluth (black squares) data of the discrete derivatives of $G_E\lt Q^2\rt$ multiplied by $(-6)$ and in units of fm$^2$, to have a direct connection with the mean square radius, see Eq.~\eqref{eq:r-der}. For illustration of the sensitivity of the charge FF to the value of the radius, two areas indicate a large interval $0.80\le R_E\le 0.92$ fm (light-blue)  and an interval close to the measured radii: $0.84\le R_E \le 0.88$ fm (blue).

%
%
Fig.~\ref{fig:delta-delta} shows the ratios of relative errors of discrete derivative and electric FF, i.e.,
\be
\frac{\delta\Delta G^{S,R}_{E,j}/\Delta G^{S,R}_{E,j}}{\delta \ov{G}^{S,R}_{E,j}/\ov{G}^{S,R}_{E,j}}
\,,\hh	j\ug1,\dots,N_{S,R}-1\,,
\nen
that, being $\ov{G}^{S,R}_{E,j}$ and $\delta\ov{G}^{S,R}_{E,j}$  mean values of the FF and their errors, i.e.,
\be
\ov{G}^{S,R}_{E,j}=\frac{G^{S,R}_{E,j+1}+G^{S,R}_{E,j}}{2}\,,\hh
\delta\ov{G}^{S,R}_{E,j}=\frac{\sqrt{\lt\delta G^{S,R}_{E,j+1}\rt^2+\lt\delta G^{S,R}_{E,j}\rt^2}}{2}
\,,\hh	j\ug1,\dots,N_{S,R}-1\,,
\nen
can be written as
\be
\frac{\delta\Delta G^{S,R}_{E,j}/\Delta G^{S,R}_{E,j}}{\delta \ov{G}^{S,R}_{E,j}/\ov{G}^{S,R}_{E,j}}=\frac{G^{S,R}_{E,j+1}+G^{S,R}_{E,j}}{\left|G^{S,R}_{E,j+1}-G^{S,R}_{E,j}\right|}\,.
\nen
\\
It follows that the more points are close, ${Q}^{2S,R}_j\to{Q}^{2S,R}_{j+1}$, the more the ratio increases, it diverges as $1/\lt {Q}^{2S,R}_{j+1}-{Q}^{2S,R}_{j}\rt$, indeed  
\be
\frac{\delta\Delta G^{S,R}_{E,j}/\Delta G^{S,R}_{E,j}}{\delta \ov{G}^{S,R}_{E,j}/\ov{G}^{S,R}_{E,j}}
\mathop{\mathlarger{\mathlarger{\sim}}}_{{Q}^{2S,R}_j\to{Q}^{2S,R}_{j+1}}
\frac{2G^{S,R}_{E,j}}
{\left|\frac{d G_{E}}{d Q^{2S,R}}(Q_j^{2S,R})\right| \lt Q^{2S,R}_{j+1}-Q^{2S,R}_{j}\rt}\,.
\nen
The ratio of spline data becomes larger than $10^2$ below $Q^2=0.02$ GeV$^2$, while the ratio of Rosenbluth data exceeds $10^2$ just for $Q^2\le 0.1$ GeV$^2$. It follows that the expected error on the extracted proton radius, which does correspond to the error on the derivative and not to that on the FF itself, turns to be larger by two orders of magnitude compared to the precision on $G_E$. This means that a relative error of a percent on the measurement of the elastic cross section translates into a few percent on the precision of the electric FF (that saturates the cross section at low $Q^2$), and finally to more than 100\% on its derivative, even in the case of the model-shaped spline data.  

Therefore, despite the fact that the analysis of cross section data based on predefined FF functional forms gives a good point by point representation for FFs themselves, such an analysis does introduce a large uncertainty in the derivative and hence in the extracted radius.
%
%
 \section{The fitting procedure}
%
We consider different fit scenarios depending on: data sets, Rosenbluth and spline; observables, discrete derivative alone and together with the FF itself; fit function, polynomials inferred by the Taylor series of various degrees; $Q^2$ intervals.
\\
Our primary result is obtained by fitting Rosenbluth data on the discrete derivative of $G_E$, for two main reasons:
\begin{itemize}
\item the discrete derivative is the observable directly related to the radius;
\item the Rosenbluth technique allows to extract FFs from elastic cross section data in model-independent way.
\end{itemize}
%
%
\begin{figure}[h!]
\begin{center}
	\includegraphics[width=.49\textwidth]{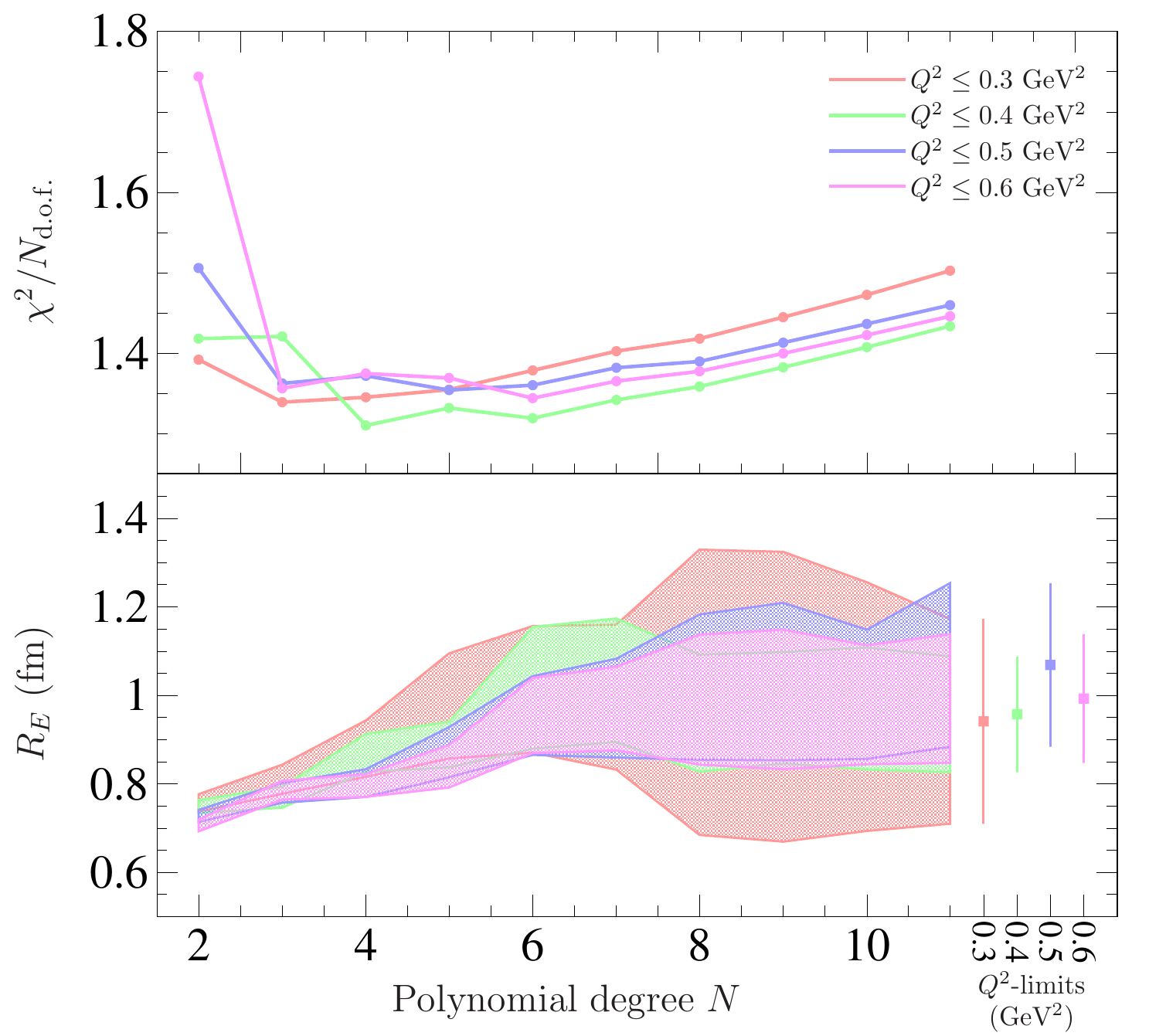}\hfill%
	\includegraphics[width=.49\textwidth]{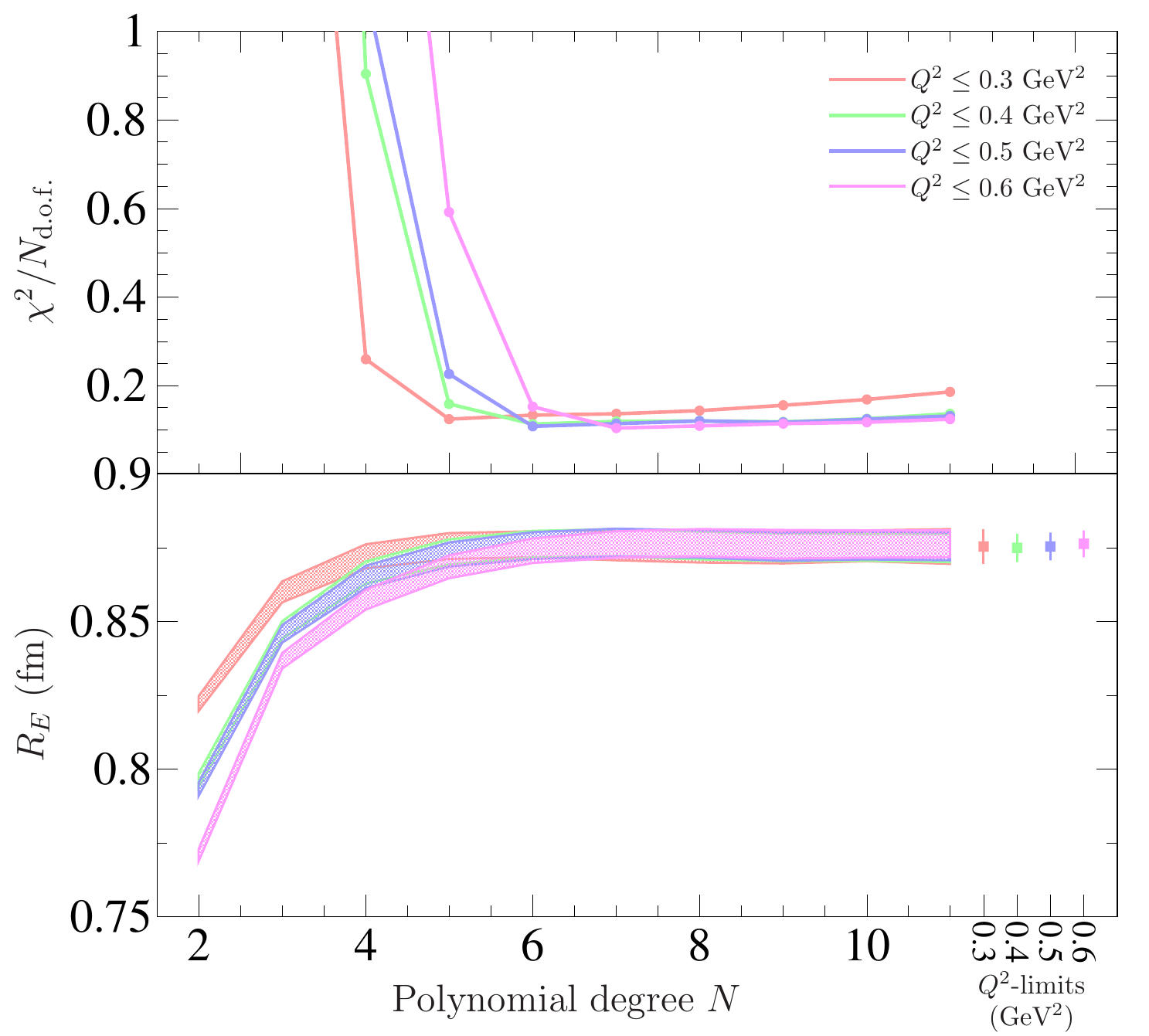}\vspace{-4mm}
		\caption{Normalized $\chi^2$ and proton radius versus the polynomial degree $N$ of fit functions defined in Eq.~\eqref{eq:fit-fun}, obtained by fitting the discrete derivative from Rosenbluth data, left panel; from spline data, right panel. The shaded areas represent errors on the extracted radii and the different colours refer to the different $Q^2$ fitting intervals. Points with error bars, in the lower panels, highlight values of the proton radius extracted considering polynomials of highest orders. \label{fig:chi2-raggio}}
\end{center}	
\end{figure}
The left panel of Fig.~\ref{fig:chi2-raggio} shows the normalized $\chi^2$'s and the extracted proton radii, obtained by fitting the discrete derivative of Rosenbluth data only, as a function of the degree of the fitting polynomial given in Eq.~\eqref{eq:fit-fun}, for the indicated four $Q^2$ intervals. Colored points with errors in the right-hand side of lower panel represent the values of the radius obtained with the highest polynomial degree as a function of the upper limit of the $Q^2$ intervals. 
\\
These values and the corresponding $\chi^2$, for the four $Q^2$ intervals, are reported in Table~\ref{tab:1}. The large errors on the extracted $R_E$ clearly demonstrate the difficulty to use data - even very precise - on a physical observable, to extract the derivative of such a quantity in a region that is not experimentally accessible. 
\\
On the other hand, the stability of the results obtained by using spline data, $\chi^2$ values and radii, that are shown in the right panel of Fig.~\ref{fig:chi2-raggio}, and reported in Table~\ref{tab:1} for the polynomial of highest degree, represents an undeniable indication of how much a model-dependent extraction of $G_E$ from cross section data does influence the result.
%
%
\begin{figure}[h!]
\begin{center}
	\includegraphics[width=.49\textwidth]{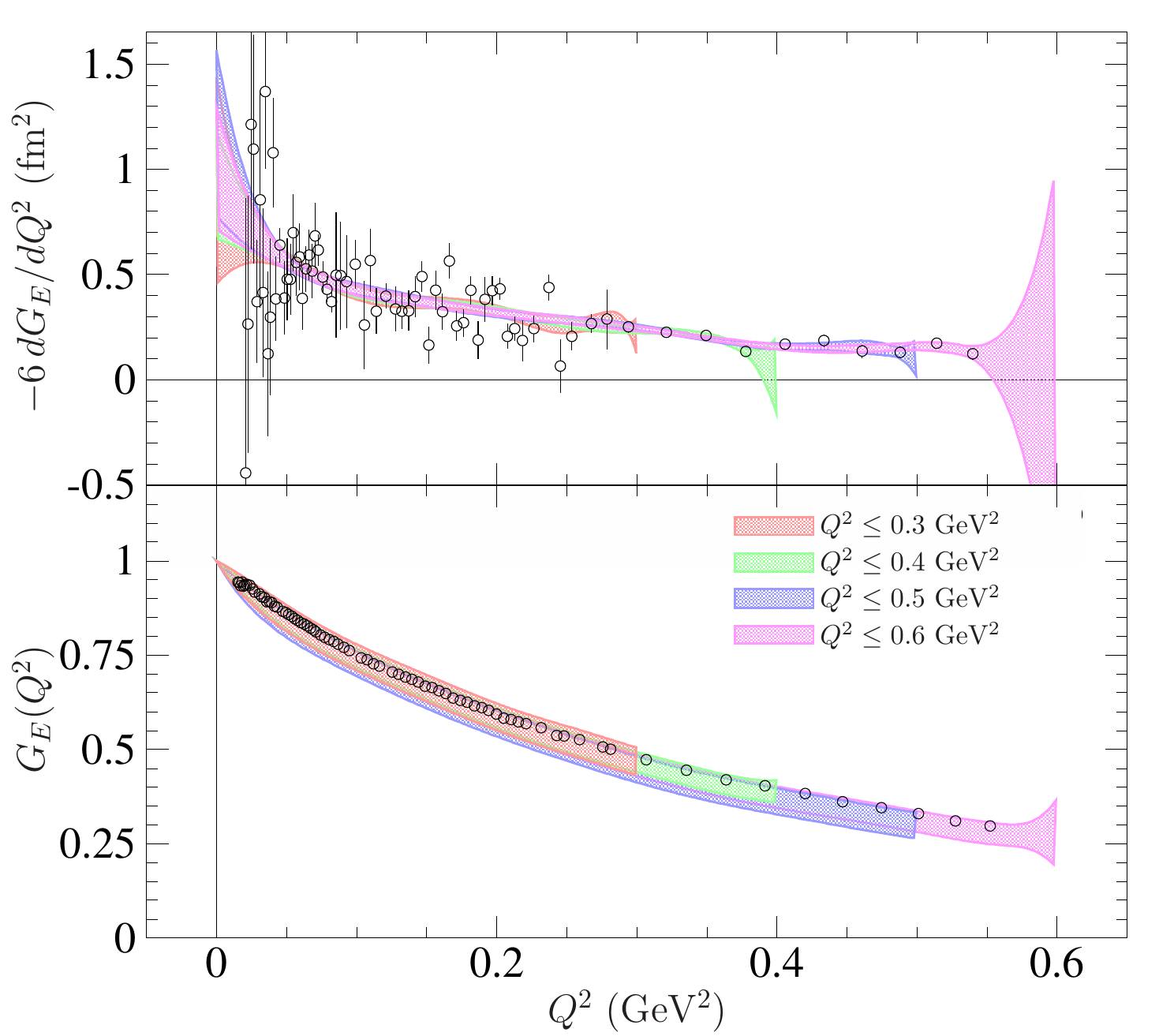}\hfill%
	\includegraphics[width=.49\textwidth]{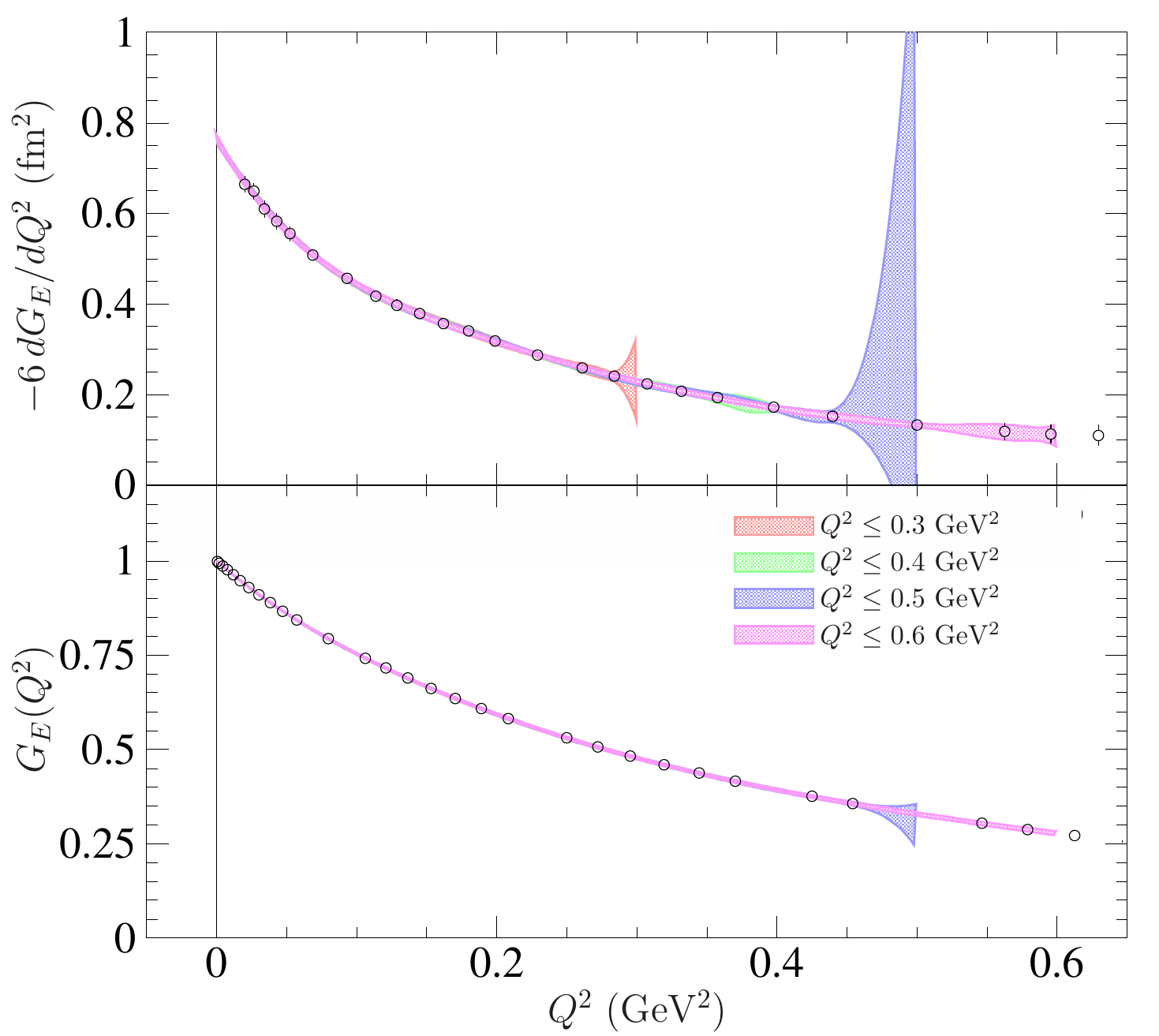}
		\caption{\label{fig:derivata-ff} Data and polynomial fit on the discrete derivative of the electric proton FF multiplied by $(-6)$ and FF itself.
		Colored curves represent fits, only on the derivative, of Rosenbluth (left panel) and spline data (right panel) for the four specified $Q^2$ intervals.}
\end{center}	
\end{figure}\\
Fig.~\ref{fig:derivata-ff} shows, in case of Rosenbluth and spline extractions, in left and right panel respectively, data and fit functions, as error bands~\footnote{\label{foot:1}Errors on parameters and hence on the fit function have been obtained by means of a Monte Carlo procedure consisting in the following steps: $N$ sets of data are generated by Gaussian fluctuations of the original ones; $N$ sets of parameters $\{p_{1,j},\ldots,p_{M,j}\}_{j=1}^N$ 
are obtained by fitting the generated data sets; best values and errors for parameter, as well as for curves, at each $Q^2$ point, are obtained by the standard statistical analysis, i.e., by taking the mean and the standard deviation
$$
\overline{p}_k=\sum_{j=1}^N \frac{p_{k,j}}{N}\,,\hspace{5mm}
\delta\overline{p}_k=\sqrt{\sum_{j=1}^N \frac{\lt p_{k,j}-\overline{p}_k\rt^2}{N-1}}\,,\hspace{5mm} k=1,\ldots,M\,.
$$
The corresponding set of fit functions is $\{f(p_{1,j},\ldots,p_{M,j};Q^2)\equiv f_j(Q^2)\}_{j=1}^N$, where $f_j(Q^2)$ stands for the fit function defined by the $j$-th parameter set $\{p_{1,j},\ldots,p_{M,j}\}_{j=1}^N$. At each $Q^2$, lower and upper edges of the error band are defined as $\overline{f}(Q^2)\pm \delta\overline{f}(Q^2)$, with
$$
\overline{f}(Q^2)=\sum_{j=1}^N \frac{f_{j}(Q^2)}{N}\,,\hspace{5mm}
\delta\overline{f}(Q^2)=\sqrt{\sum_{j=1}^N \frac{\lt f_{j}(Q^2)-\overline{f}(Q^2)\rt^2}{N-1}}\,.
$$ 
}, on electric FF and its derivative, in units of fm$^2$ and multiplied by $(-6)$. The fit has been performed exploiting the second $Q^2$ polynomial of Eq.~\eqref{eq:fit-fun}, in the four specified $Q^2$ intervals, only on the discrete derivative. The obtained parameters $R_E^2$, $\Delta_2^2$,\dots, $\Delta^2_{11}$, are then used in the first $Q^2$ polynomial of the same Eq.~\eqref{eq:fit-fun} to compute the electric FF, assuming the normalization $G_E(0)=1$. Corresponding error bands are shown superimposed to the data in the lower panels of Fig.~\ref{fig:derivata-ff}. 
\\
Fig.~\ref{fig:chi2-raggio2} shows normalized $\chi^2$'s and extracted proton radii, in case of Rosenbluth, left panel, and spline data, right panel, obtained by a combined fit of discrete derivative and FF data, performed by using the polynomials of Eq.~\eqref{eq:fit-fun}. The obtained error bands are shown in Fig.~\ref{fig:derivata-ff2} superimposed to the data. 
%
%
\begin{figure}[h!]
\begin{center}
	\includegraphics[width=.49\textwidth]{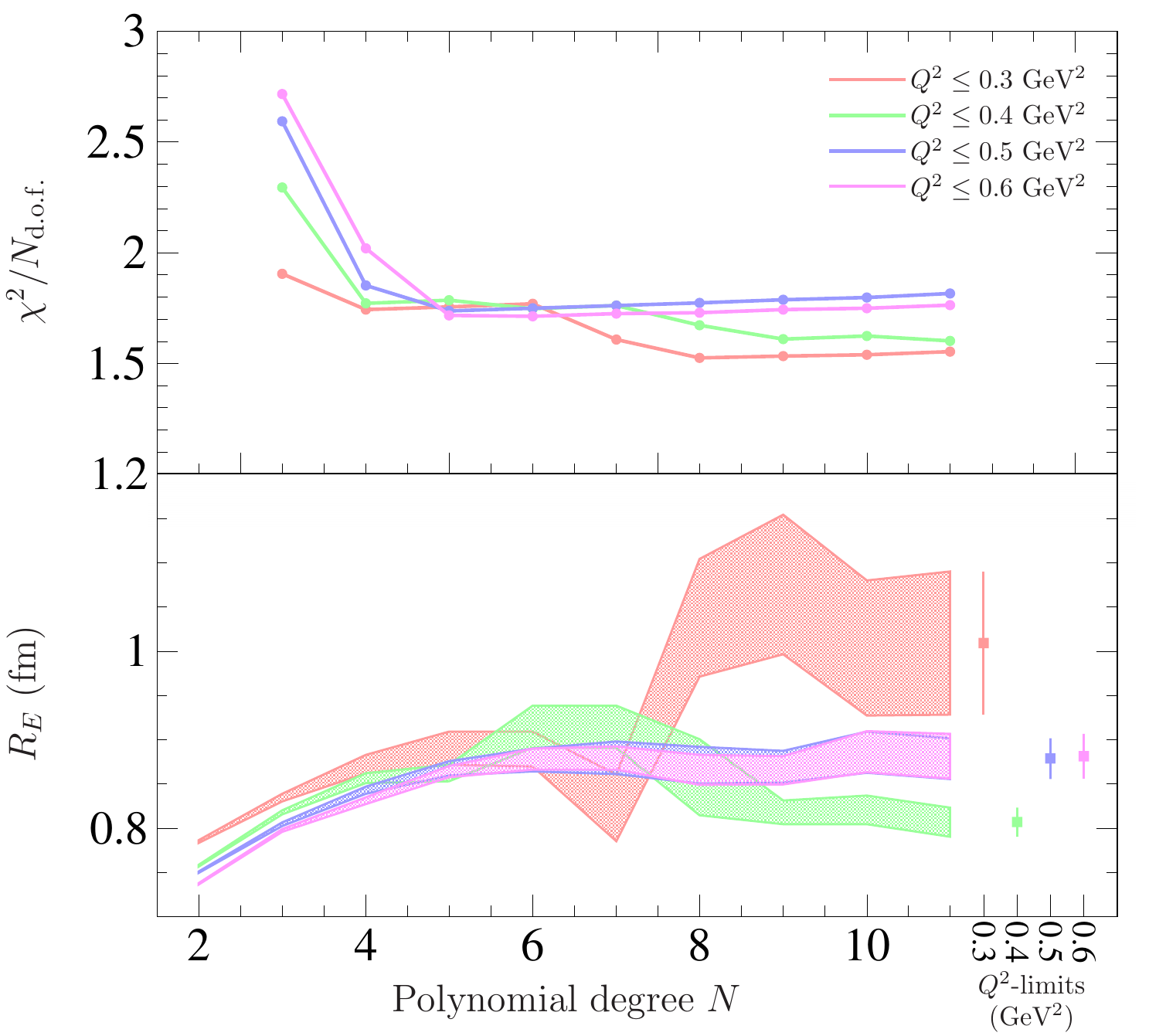}\hfill%
	\includegraphics[width=.49\textwidth]{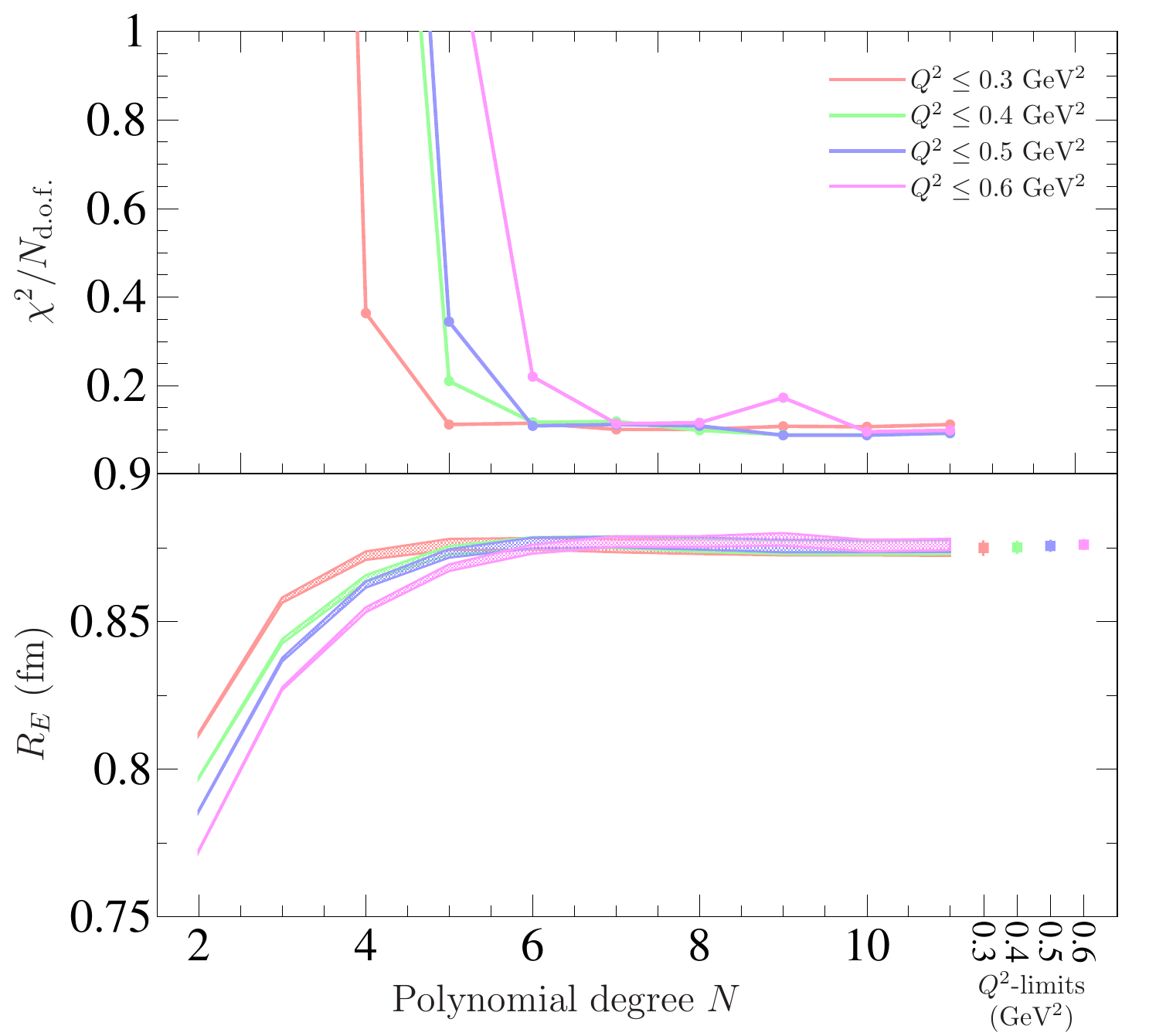}\vspace{-4mm}
		\caption{Normalized $\chi^2$ and proton radius versus the polynomial degree $N$ of fit functions defined in Eq.~\eqref{eq:fit-fun}, obtained by fitting the discrete derivative from Rosenbluth data, left panel; from spline data, right panel. The shaded areas represent errors on the extracted radii and the different colours refer to the different $Q^2$ fitting intervals. Points with error bars, in the lower panels, highlight values of the proton radius extracted considering polynomials of highest.\label{fig:chi2-raggio2}}
\end{center}	
\end{figure}\\
As reported in Table~\ref{tab:1}, the additional constraint of FF data reduces the error on the extracted radii by a factor of $5-10$ in case of Rosenbluth data: from $\sim15\%$ to $\sim 3\%$, for largest $Q^2$ interval; and a factor of two for spline data: from $\sim0.4\%$ to $\sim 0.2\%$, for largest $Q^2$ interval.
\\
However, it does remain a discrepancy between the results for the radius of the two data sets. Moreover, while the radii extracted from spline data are very stable against the different fit procedures, those obtained by using Rosenbluth data show a dependence on the polynomial degree, the $Q^2$ interval and the inclusion of the data set on the FF, in addition to that on the discrete derivative.
\\
This represents a further demonstration of both, the model-dependence inherent in the spline data and the loss of accuracy in passing from cross section to the FF derivative. It is this unavoidable loss of accuracy that does prevent the possibility of reaching a \textperthousand\ precision in the radius values.      
%
%
\begin{figure}[h!]
\begin{center}
	\includegraphics[width=.49\textwidth]{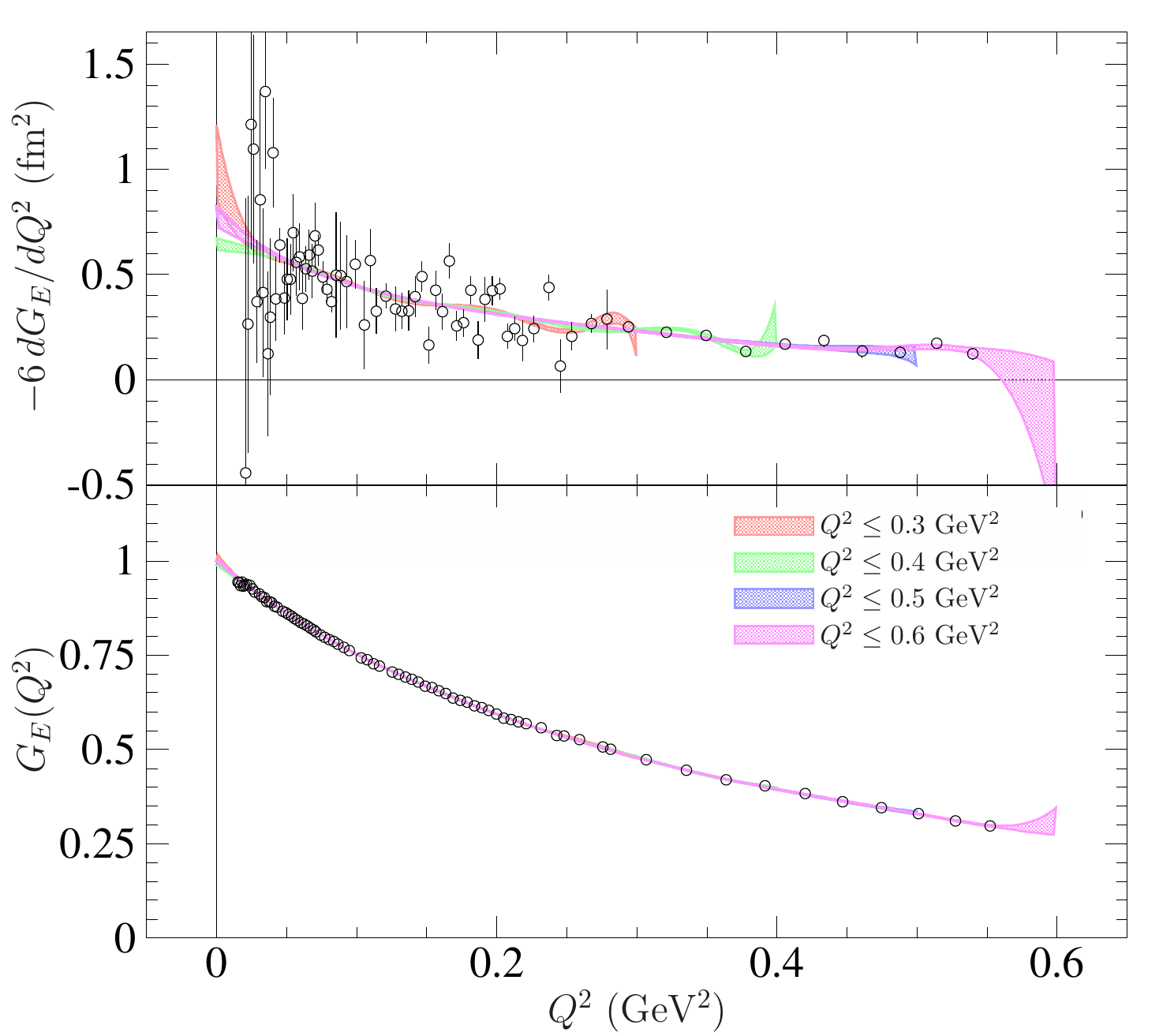}\hfill%
	\includegraphics[width=.49\textwidth]{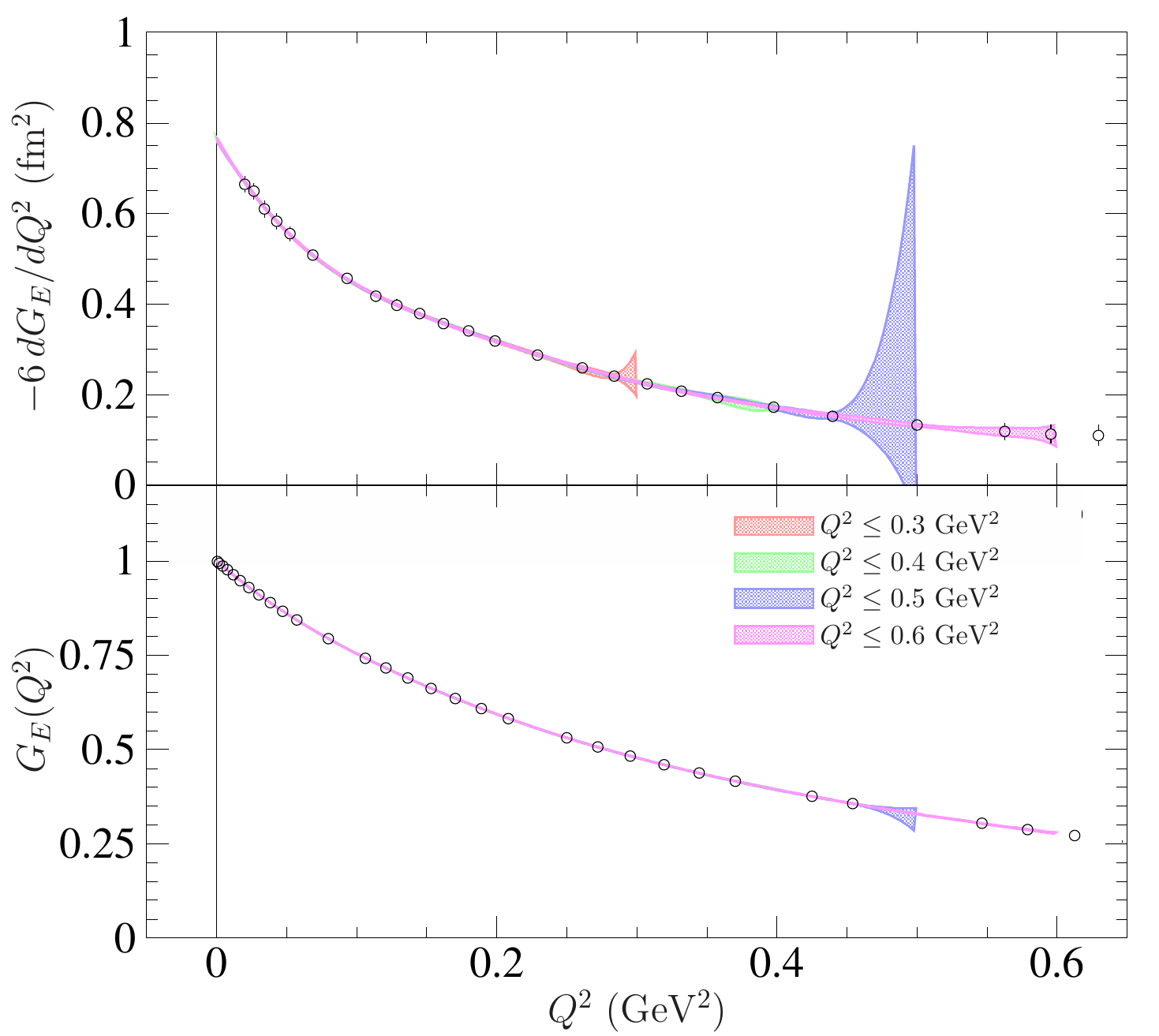}
		\caption{\label{fig:derivata-ff2} Data and polynomial fit on the discrete derivative of the electric proton FF multiplied by $(-6)$ and FF itself.
		Colored curves represent fits, only on the derivative, 
		of Rosenbluth, left panel, and spline data, right panel, in the four specified $Q^2$ intervals.}
\end{center}	
\end{figure}\\
\begin{table}[h!]
\begin{center}
\caption{\label{tab:1} Normalized $\chi^2$'s and radii for the studied data combinations and fitting procedures at the highest considered degree, $N=11$.}\vspace{0mm}
		\begin{tabular}{ c | c || c | c || c | c }
			\multicolumn{2}{c||}{} & \multicolumn{2}{|c||}{Rosenbluth} & \multicolumn{2}{c}{Spline} \\ 
\cline{3-6}
\multicolumn{2}{c||}{} & $\chi^2/N_{\rm d.o.f.}$ & $R_E$ (fm) & $\chi^2/N_{\rm d.o.f.}$ & $R_E$ (fm)\\
\hline\hline
\multirow{2}{*}{$Q^2\le 0.3$ GeV$^2$} & $dG_E/dQ^2$ &
1.50 & $0.9411 \pm 0.2310$ & 0.19 & $0.8754 \pm 0.0059$ \\ 
 \cline{2-6} & $G_E\cup dG_E/dQ^2$ &
1.55 & $1.0088 \pm 0.0809$ & 0.11 & $0.8749 \pm 0.0026$ \\ 
 \hline\hline
\multirow{2}{*}{$Q^2\le 0.4$ GeV$^2$} & $dG_E/dQ^2$ &
1.43 & $0.9568 \pm 0.1309$ & 0.14 & $0.8749 \pm 0.0048$ \\ 
 \cline{2-6} & $G_E\cup dG_E/dQ^2$ &
1.60 & $0.8070 \pm 0.0164$ & 0.09 & $0.8751 \pm 0.0023$ \\ 
 \hline\hline
\multirow{2}{*}{$Q^2\le 0.5$ GeV$^2$} & $dG_E/dQ^2$ &
1.46 & $1.0681 \pm 0.1848$ & 0.13 & $0.8754 \pm 0.0047$ \\ 
 \cline{2-6} & $G_E\cup dG_E/dQ^2$ &
1.82 & $0.8786 \pm 0.0229$ & 0.09 & $0.8756 \pm 0.0020$ \\ 
 \hline\hline
\multirow{2}{*}{$Q^2\le 0.6$ GeV$^2$} & $dG_E/dQ^2$ &
1.45 & $0.9927 \pm 0.1453$ & 0.12 & $0.8763 \pm 0.0046$ \\ 
 \cline{2-6} & $G_E\cup dG_E/dQ^2$ &
1.76 & $0.8811 \pm 0.0253$ & 0.10 & $0.8761 \pm 0.0019$ \\ 
%
		\end{tabular}
	\end{center}
\end{table}\\
Let us considered the fit for $Q^2\le 0.6$ GeV$^2$. Depending on data sets and observables, four values of the radius are extracted, see Table~\ref{tab:1},
\be
\begin{array}{rcl c rcl}
 R_E^{R,1C}
\ug 0.99\pm 0.15\,{\rm fm}\,, &\hh& 
 R_E^{S,1C}
 \ug 0.876\pm 0.005\,{\rm fm}\,,\\
 && && &&\\
 R_E^{R,2C}
 \ug 0.88\pm 0.03\,{\rm fm}\,, &\hh& 
  R_E^{S,2C}
  \ug 0.876\pm 0.002\,{\rm fm}\,,\\
\end{array}
\nen  
where the apex stands for data sets: "$R$" indicates Rosenbluth and "$S$" spline, same notation of Sec.~\ref{sec:data}; and observables: $1C$ means that the minimization is performed only on the derivative, one constraint, and $2C$ on derivative and FF, two constraints.
\\
The main result, $R_E^{R,1C}$, is the one obtained by fitting the Rosenbluth data directly on the discrete derivative, its large error, $\sim 15\%$, reflects the accuracy of the data on the discrete derivative, see right panel of Fig.~\ref{fig:data-ge-dge}. The inclusion  in the minimization procedure of FF data enhances the accuracy on the radius and changes the central value, which is reduced of about $12\%$. Nevertheless, due to the large errors, the two results remain compatible.
\\
%
The spread of these results, summarized in Table~\ref{tab:1}, leads us to infer that elastic electron-proton scattering does not represent the best method to extract the proton radius. The difference of $\sim 12\%$ among the values extracted from the same data on the cross section measurements  should be considered as the size of the systematic errors associated to the method.  Attributing  such an error to the radius extracted from elastic scattering data brings into agreement all measurements, in atomic and hadron physics.

\section{Conclusion}
%
This paper has the aim to  drive the attention of the reader to the relation between the precision of the measured observable and the precision on the physical quantity of interest. Large efforts are undertaken in  measuring the cross section at the lowest $Q^2$ with the highest precision. We showed that, when the quantity that we want to determine, in the present case the proton radius, is  related to the derivative of the measured observable, the precision on the quantity  itself becomes worse by orders of magnitude.
\\
This precision can be artificially reduced, if one adds physical constrains or another input, as a predefined function that the (unknown) quantity should follow. In this case, care should be taken in attributing a "theoretical" or "systematical" error to the results.  
 \\
Moreover, when an extrapolation is needed as the experiment cannot reach the kinematic region of interest, 
 the introduction of pre-defined functional forms for the investigated observables does constrain the limit of the derivative. In this sense, the direct extrapolation of the discrete derivative of the FF itself is more correct, although, of course one does not take advantage of the whole set of measurements and looses precision. 
 
\begin{figure}[h!]
\begin{center}
	\includegraphics[width=.49\textwidth]{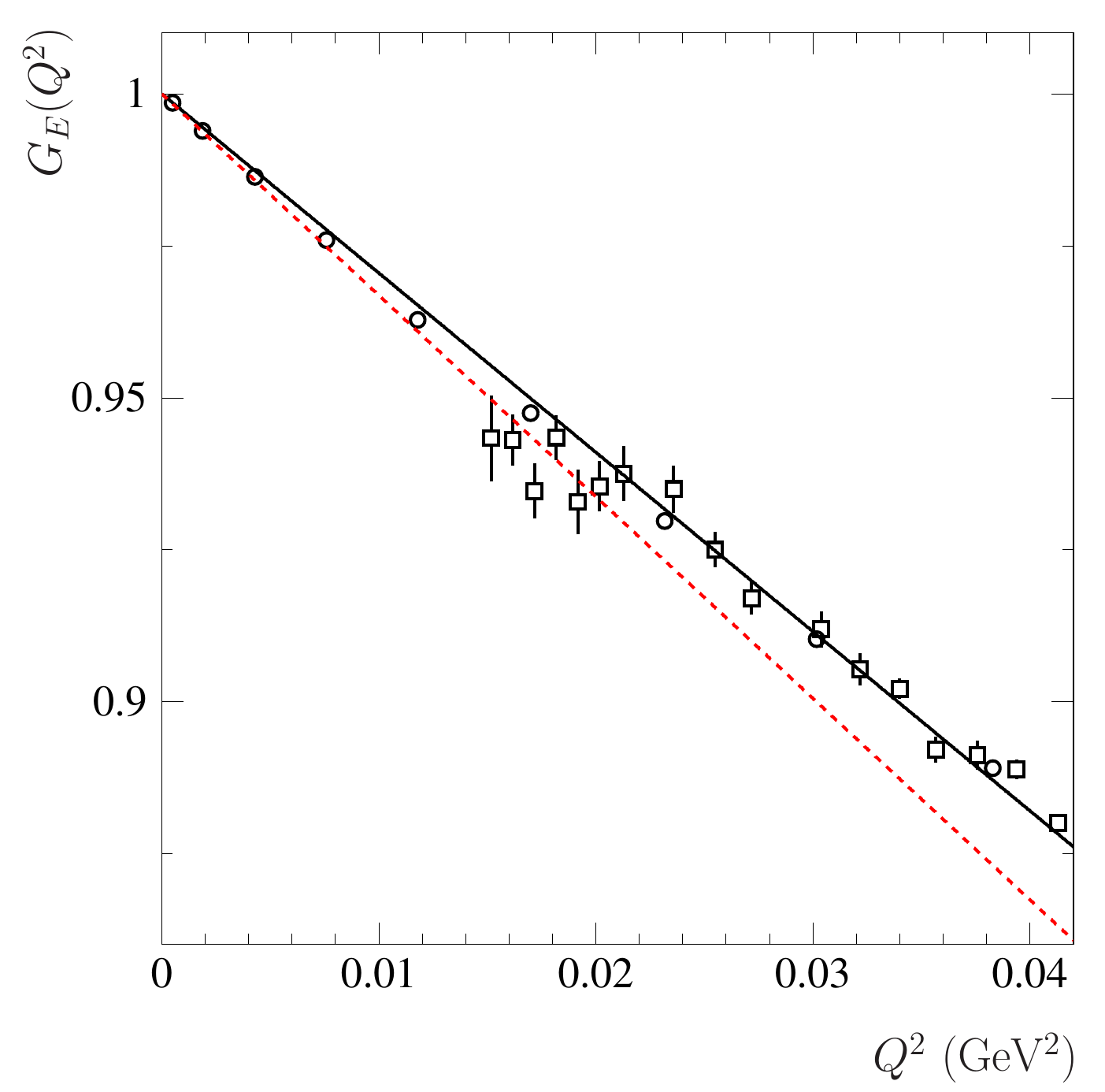}
		\caption{
\label{Fig:Figslopes} 
Data from  spline (open circles) and Rosenbluth separation (open squares) from Ref. \cite{Bernauer:2010wm}, and constrained first order polynomials with slopes corresponding to two values of the radius: $R_E=0.83$ fm (black solid line) and $R_E=0.87$ fm (red dashed line).}
\end{center}	
\end{figure}

In order to give a simple and visual illustration, in Fig. \ref{Fig:Figslopes} the low-$Q^2$ data points from spline and Rosenbluth methods, Ref. \cite{Bernauer:2010wm},  are shown together with two first order polynomials  constrained to unity at $Q^2=0$, with slopes corresponding to the two 'nominal' values of the radius : $R_E=0.83$ fm (black solid line) and $R_E=0.87$ fm (red dashed line). The smaller slope is closer to the data. This figure shows the effect of high order terms, of the extrapolation  as well as of the need of a free normalization at $Q^2=0$ that are necessary to reach a good fit.
\\
\\

Let us stress that, in the game of reaching the lowest possible $Q^2$, the situation of inverse kinematics,  i.e., the elastic reaction when a proton beam hits an electron target, allows to reach $Q^2$ values smaller by several orders of magnitude~\cite{Gakh:2013pda}. If this may help, due to the fact that the extrapolation of the cross section is reduced to a smaller interval, it does not solve the problems pointed out above: the transition from the physics of a scattering problem to a compound object, of the spread of the derivatives, as well as the necessity to introduce model dependent effects, as Coulomb corrections.
\\
The conclusion of this work is that the extraction of  the radius, a static quantity,  from a dynamical object as the cross section, is by construction affected by large systematics that cannot be reduced by the intrinsic nature of the measurement. 
\section{Acknowledgments}
Thanks are due to the colleagues and participants of the 668.WE-Heraeus-Seminar on "Baryon Form Factors: where do you stand", in particular to Rinaldo Baldini Ferroli, Alaa Dbeyssi, Eric Voutier and Alexander~Milstein for useful discussions and constructive remarks.  We acknowledge Mehdi Drissi for interesting exchanges on mathematical methods.
\clearpage
\begin{widetext} 
\revappendix
\section{The Schwarz's inequality for mean values of even powers of the radius}
\label{app:Schwarz}
For any pair functions $f(x)$, $g(x)\in L^2(E)$, where $L^2(E)$ represents the set of functions which are quadratically Lebesgue integrable on $E\subset\R$, it holds the Schwarz's inequality~\cite{Schwarz:1885}
\be
\left|\int_E f^*(x)g(x)dx\right|^2
\le
\int_E \left|f(x)\right|^2dx \int_E \left|g(x)\right|^2dx\,.
\label{eq:SI}
\en
Given a spherically symmetric density $\rho\lt\vec{r}\rt\equiv\rho(r)$, the mean value of $r^n$, i.e., the normalized $n$-th moment of the density around the origin, with $n\in\N$, is
\be
\left\langle r^n\right\rangle\ug 
\frac{\ds\int  r^n\rho(r)  d^3\vec{r}}{\ds\int  \rho(r)  d^3\vec{r}}=
\frac{\ds4\pi\int_0^\infty r^{n+2}\rho(r) \, dr}{\ds4\pi\int_0^\infty r^{2} \rho(r) \, dr}\,.
\nen
By defining the radial density $\eta(r)=4\pi r^2\rho(r)$, and assuming the unitary normalization
\be
\int  \rho(r)  d^3\vec{r}=
4\pi\int_0^\infty r^{2} \rho(r) \, dr=
\int_o^\infty\eta(r)dr=
1\,,
\nen
the $n$-th moment around $r=0$ can be written as
\be
\left\langle r^n\right\rangle=\int_0^\infty r^n\eta(r) \, dr\,.
\nen
If the mean values of all integer powers of the radius, or, equivalently, all moments around the origin of the radial density $\eta(r)$ exist and are limited, then the function $f_n(r)=\sqrt{\eta(r)}\,r^n \in L^2(0,\infty)$, for all $n\in\N$, indeed
\be
\int_0^\infty\left|f_n(r)\right|^2dr=
\int_0^\infty r^{2n}\eta(r)dr=\left\langle r^{2n}\right\rangle<\infty\,.
\nen
Using the Schwarz's inequality of Eq.~\eqref{eq:SI}, with $f(r)=f_n(r)=\sqrt{\eta(r)}\,r^n$ and $g(r)=f_0(r)=\sqrt{\eta(r)}$, we obtain the inequality $\left\langle r^n \right\rangle^2\le\left\langle r^{2n}\right\rangle$, indeed 
\be
\left|\int_0^\infty\!\!\!\! f^*(r)g(r)dr\right|^2
&\le&
\int_0^\infty\!\!\!\! |f(r)|^2 dr
\int_0^\infty\!\!\!\! |g(r)|^2 dr
\no\\
\lt \int_0^\infty\!\!\!\!\eta(r)\,r^n dr\rt^2
&\le&
\int_0^\infty\!\!\!\! \eta(r)\,r^{2n} dr 
\int_0^\infty\!\!\!\! \eta(r)dr
\no\\
\left\langle r^n \right\rangle^2
&\le&
\left\langle r^{2n}\right\rangle\,.
\label{eq:even}
\en 
On the other hand, with $f(r)=\sqrt{\eta(r)}\,r^{n-1/2}$ and $g(r)=\sqrt{\eta(r)}\,r^{1/2}$, using Eq.~(\ref{eq:even}) with $n=1$, i.e., $\left\langle r\right\rangle\le \left\langle r^2\right\rangle^{1/2}$,
\be
\lt \int_0^\infty\eta(r)\,r^n\, dr\rt^2
&\le&
\int_0^\infty \eta(r)\,r^{2n-1} dr \int_0^\infty \eta(r)\,r \,dr
\no\\
\left\langle r^n \right\rangle^2
&\le&
\left\langle r^{2n-1}\right\rangle
\left\langle r\right\rangle
\le
\left\langle r^{2n-1}\right\rangle
\left\langle r^2\right\rangle^{1/2}
\,,
\nen 
so that
\be
\left\langle r^{2n-1}\right\rangle\ge
\left\langle r^n\right\rangle^2
\left\langle r^2\right\rangle^{-1/2}\,.
\label{eq:odd}
\en
Inequalities of Eqs.~\eqref{eq:even} and~\eqref{eq:odd} can be generalized as it follows
\be
\left\langle r^{2n-b}\right\rangle\ge
\left\langle r^n\right\rangle^2
\left\langle r^2\right\rangle^{-b/2}\,,\hh \forall\,n\in\N\,,
\label{eq:dis-gen}
\en
with $b\in\{0,1\}$. In order to apply iteratively the inequality of Eq.~\eqref{eq:dis-gen}, we consider, for a generic integer $n\in\N$, the following expression in terms of powers of two
\be
n=2^N-\sum_{j=0}^{N-2}b_j 2^{j}
=2^N-\sum_{j=1}^{N-2}b_j 2^{j}-b_0
=2\lt 2^{N-1}-\sum_{j=1}^{N-2}b_j 2^{j-1}\rt-b_0\,,
\label{eq:n-def}
\en 
where
\be
N=\min_{j\in\N}\left\{2^j\ge n\right\}\,,
\label{eq:N-def}
\en 
and the $(N-1)$-tuple $\lt b_0,b_1,\ldots,b_{N-2}\rt\in\left\{ 0,1\right\}^{N-1}$. It is quite easy to show that such an expression always exists, indeed the sequence of 0's and 1's, represented by the $(N-1)$-tuple $\lt b_0,b_1,\ldots,b_{N-2}\rt$, is the difference $2^N-n$ written in base-2 (binary notation), i.e.,
\be
2^N-n=\lt b_{N-2}b_{N-3}\ldots b_{1}b_{0}\rt_2\,,
\label{eq:diff-def}
\en
(the number "2" subscript on the parentheses indicates the base-2 notation).
In particular it does contain only $N-1$ digits, because the difference between $2^N$ and a generic binary number with $N$ digits is lower than $2^{N-1}$. In fact, if the difference of Eq.~\eqref{eq:diff-def} would be a binary number with $N-1$ digits $\lt b_{N-1}b_{N-2}\ldots b_{1}b_{0}\rt_2$, where $b_{N-1}=1$, then $\lt b_{N-1}b_{N-2}\ldots b_{1}b_{0}\rt_2\ge 2^{N-1}$ and from Eq.~\eqref{eq:diff-def}
\be
n=2^N-\lt b_{N-1}b_{N-2}\ldots b_{1}b_{0}\rt_2
\le 2^N-2^{N-1}=2^{N-1}\,.
\nen
But this is in contradiction with the definition of $N$ given in Eq.~\eqref{eq:N-def}, from which we would arrive at the absurd
\be
N=\min_{j\in\N}\left\{2^j\ge n\right\}=N-1\,.
\nen 
Consider the sequence of inequalities that can be obtained using the inequality of Eq.~\eqref{eq:dis-gen} and the expression of $n$ of Eq.~\eqref{eq:n-def}
\be
\la r^{2n}\ra
& \ge & 
\la r^n\ra^2
=\la r^{2\lt 2^{N-1}-\sum_{j=1}^{N-2}b_j 2^{j-1}\rt-b_0}\ra^2
\no\\
& \ge & 
\la r^{2^{N-1}-\sum_{j=1}^{N-1}b_j 2^{j-1}}\ra^{2^2}
\la r^2\ra^{-b_0}
=\la r^{2\lt2^{N-2}-\sum_{j=2}^{N-2}b_j 2^{j-2}\rt-b_1}\ra^{2^2}
\la r^2\ra^{-b_0}
\no\\
& \ge & 
\la r^{2^{N-2}-\sum_{j=2}^{N-2}b_j 2^{j-2}}\ra^{2^3}
\la r^2\ra^{-b_0-2b_1}
=\la r^{2\lt 2^{N-3}-\sum_{j=3}^{N-2}b_j 2^{j-3}\rt-b_2 }\ra^{2^3}
\la r^2\ra^{-b_0-2b_1}
\no\\
& \ge & 
\la r^{2^{N-3}-\sum_{j=3}^{N-2}b_j 2^{j-3}}\ra^{2^4}
\la r^2\ra^{-b_0-2b_1-2^2b_2}
=\la r^{2\lt 2^{N-4}-\sum_{j=4}^{N-2}b_j 2^{j-4}\rt-b_3 }\ra^{2^4}
\la r^2\ra^{-b_0-2b_1-2^2b_2}
\no\\
& \ge & 
\cdots
\no\\
& \ge & 
\la r^{2^{N-k}-\sum_{j=k}^{N-2}b_j 2^{j-k}}\ra^{2^{k+1}}
\la r^2\ra^{-\sum_{j=0}^{k-1}b_j 2^j}
\no\\
& \ge & 
 \cdots
\no\\
& \ge & 
\la r^{2^{2}-\sum_{j=N-2}^{N-2}b_j 2^{j-N+2}}\ra^{2^{N-1}}
\la r^2\ra^{-\sum_{j=0}^{N-3}b_j 2^j}
=\la r^{2^{2}-b_{N-2}}\ra^{2^{N-1}}
\la r^2\ra^{-\sum_{j=0}^{N-3}b_j 2^j}
\,,\hh k=N-2\,,
\nen
where the last one is the $(N-2)$-th iteration.
From Eq.~\eqref{eq:dis-gen} follows that the left-hand-side of the last inequality is greater than $\la r^{2}\ra^{n}$, indeed 
\be
\begin{array}{rc l cl}
\la r^{2n}\ra
& \ge &
\la r^{2^{2}-b_{N-2}}\ra^{2^{N-1}}
\la r^2\ra^{-\sum_{j=0}^{N-3}b_j 2^j} &&
\no\\
& \ge &
\lt\la r^{2}\ra^2 \la r^{2}\ra^{-b_{N-2}/2}\rt^{2^{N-1}}
\la r^2\ra^{-\sum_{j=0}^{N-3}b_j 2^j}
\ug
\la r^{2}\ra^{2^N-b_{N-2}2^{N-2}-\sum_{j=0}^{N-3}b_j 2^j} \\
&&
\ug
\la r^{2}\ra^{2^N-\sum_{j=0}^{N-2}b_j 2^j}
=
\la r^{2}\ra^{n}\,,\\
\end{array}
\nen
where the last identity follows from Eq.~\eqref{eq:n-def}, in summary we obtain the inequality
\be
\la r^{2n}\ra\ge \la r^{2}\ra^{n}\,.
\label{eq:general}
\en
\end{widetext} 
\bibliography{Biblio}
\end{document}